\documentclass[11pt]{article}
\usepackage{amsmath,amsfonts,amssymb}
\usepackage{graphicx,xcolor}
\usepackage{geometry}
\usepackage[hidelinks]{hyperref}
\usepackage{authblk}
\usepackage{dutchcal}
 \usepackage{array,booktabs,tabularx}

\newcommand{\dif}{{\mathrm{d}}}
\newif\ifshowrevisions
\showrevisionsfalse
\newenvironment{revision}{\par\begingroup\ifshowrevisions\color{red}\fi}{\par\endgroup}
\newenvironment{newtext}{\par\begingroup}{\par\endgroup}
\newcommand{\bcdot}{\boldsymbol{\cdot}}

\newcommand{\red}[1]{#1}
\newcommand{\m}{{\mathcal m}}

\title{Coupled stochastic variational principles for multiscale surface gravity waves -- Part I: theoretical framework}
\date{}
\author[1,2]{A. Debussche}
\author[3]{E. M\'emin}

\affil[1]{Univ Rennes, CNRS, IRMAR-UMR 6625, F-35000 Rennes, France}
\affil[2]{Institut universitaire de France (IUF)}
\affil[3]{Univ Rennes, Inria, Odyssey, IRMAR-UMR 6625, Centre Inria Rennes, 35042 Rennes Cedex, France}

\begin{document}
\maketitle
\begin{abstract}

This paper develops a  stochastic variational framework for surface gravity waves. 
Starting from Luke's variational principle for irrotational, incompressible free-surface flow, 
we introduce a decomposition of the velocity potential into a resolved component and a
regularized stochastic term representing unresolved scales.
A path-wise variational principle yields the stochastic counterparts of the Laplace, Bernoulli, 
and kinematic boundary conditions. The evolution of the noise
correlation modes is derived from a linearized Euler equation, and we show that it is consistent
with a variational principle in expectation. The resulting coupled system preserves the Hamiltonian 
structure of the Zakharov--Craig--Sulem formulation. We then derive stochastic Airy, Saint-Venant, Serre--Green--Naghdi and Boussinesq approximations.
A geometric-optics analysis shows that, at
leading order, the correlation modes are transported and refracted by the resolved surface flow.
\red{Simulations of the stochastic Saint-Venant system, compared with a filtered Serre--Green--Naghdi reference, illustrate the transport and deformation of the modes and a partial correction of the resolved elevation statistics.}
This framework provides a consistent basis for reduced stochastic models of ocean waves and
for the kinetic theory developed in the companion paper.

\end{abstract}

\section{Introduction}

Waves are ubiquitous in ocean dynamics and play a central role in air-sea interaction, coastal protection, marine navigation, and climate system dynamics \cite{komen1994,young1999}. Wave models are essential for hazard monitoring associated with wave submersion, coastal erosion, and tsunamis \cite{goda2000}. While important insights can be derived from linear wave theory \cite{Airy1845}, most wave systems are intrinsically nonlinear due to the coupling between surface elevation and fluid velocity, the occurrence of breaking, and the transfer of energy across scales \cite{Lannes-2013,Zakharov68}. Nonlinear effects manifest in various forms, such as nonlinear dispersion relations, amplitude-dependent wave speeds, and energy transfers among different components of the wave system \cite{hasselmann1962nonlinear,hasselmann1963nonlinear}. However, fully nonlinear wave models, whether based on the Euler equations or on the Zakharov--Craig--Sulem Hamiltonian formulation \cite{Craig-Sulem93,Zakharov68}, remain challenging to solve numerically, especially when broadband spectra and long-time integrations are required \cite{Lannes-2013}. Moreover, surface waves constitute a vast energy reservoir \cite{Ferrari-Wunsch-09} that is only weakly and indirectly represented in models, owing to the difficulty of coupling waves and currents in numerical models \cite{Sullivan2010DynamicsWinds,VillasBoas2020WaveCurrent}.

Statistical spectral wave models, which describe the evolution of the wave action spectrum and associated statistical moments, usually referred to as phase-averaged spectral wave models, are often more tractable and computationally efficient. However, they do not accurately represent phase evolution and rely heavily on semi-empirical parameterizations of nonlinear source terms \cite{hasselmann1973,hasselmann1976,komen1994}. The classical Hasselmann kinetic equation \cite{hasselmann1962nonlinear}, derived from weak turbulence theory, has become the cornerstone of operational wave forecasting \cite{WAMDI1988,Tolman2009}. Yet this model assumes random phases and weak nonlinearity, and its numerical implementation requires careful treatment of the Boltzmann collision integral \cite{Polnikov2004}. Moreover, the role of unresolved scales, such as internal waves, submesoscale eddies, and background turbulence, in redistributing wave energy remains poorly understood and is largely absent from operational models \cite{McWilliams2016,Thomas2020}.

Here, we aim to derive stochastic dynamical models for wave evolution that bridge the gap between phase-resolved and phase-averaged descriptions. In such models, waves are represented as stochastic processes combining a phase-resolved component with a noise term associated with unresolved wave components. In this sense, the approach reconciles phase-averaged and phase-resolved descriptions, offering a unified framework in which the effects of unresolved scales are represented through a systematic stochastic closure. The formalism is fully stochastic and introduces stochastic partial differential equations to describe wave evolution, thereby providing a principled alternative to empirical source-term parameterizations. This framework builds upon recent advances in stochastic ocean dynamics modeling \cite{Bauer-et-al-JPO-20,Li-et-al-JPO-25,memin2014fluid,Resseguier2017a,Resseguier2017b,Tucciarone-et-al-James-25} and stochastic variational principles for geophysical flows \cite{cotter2017stochastic,Debussche-Memin25,Holm2015,Street-Crisan-23}. These approaches, incorporating small-scale noise advection, often referred to as transport noise, are supported by extensive recent mathematical analysis -- see, for example, \cite{Agresti-et-al-2022,Brzezniak-Slavik-2021,Carigi-Luongo-2023,crisan2019solution,Debusshe-Hug-Memin-2023,Moneyron-2025,Debussche-Pappalattera2023,Flandoli-Galeati-Luo-2021,Flandoli-Luo-2021,Flandoli-Pappalettera-2023,Flandoli-Russo2023,Galeati-Luo-2020,Galeati-Luo2023,Goodair-et-al2022,Lang2023} and references therein. 

In contrast to stochastic parameterizations involving ad hoc additive or multiplicative forcing, fluctuation-dissipation balance, together with noise-induced advection consistent with Generalized Lagrangian Mean (GLM) theory \cite{AndrewsMcIntyre1978GLM}, are naturally embedded within this formalism \cite{Bauer-et-al-JPO-20}. The introduction of time-correlated noise models and multiscale variational principles \cite{Debussche-Memin25} has recently helped connect small-scale dynamics driven by large-scale processes to Kraichnan-Tennekes random sweeping assumption \cite{Chen-Kraichnan-1989,Kraichnan-1964,Tennekes-1975}, as well as to the seminal multiscale ocean-atmosphere stochastic parameterization proposed by K. Hasselmann \cite{Frankignoul-Hasselmann-1977,Hasselmann-1976} and MTV semi-empirical approaches \cite{Majda-et-al-1999,Majda-et-al-2001}. The introduction of regularized noise processes for wave modeling also provides a methodological framework that is much easier to handle than directly considering delta-correlated noise processes \cite{dinvay2022}, which is particularly advantageous in the wave context.

\begin{revision}
Stochastic Hamiltonian formulations of surface waves have already been proposed by Dinvay and M\'emin \cite{dinvay2022} and Street \cite{Street-2023}; LU coastal-wave models, including stochastic shallow-water and dispersive approximations, were derived by Debussche, M\'emin and Moneyron \cite{debussche2024derivation} and LU shallow water model was proposed in \cite{BoulvardMeminSainteMarie2026}. However, these approaches considered prescribed noises, specified from high resolution simulations or data. The present contribution combines finite temporal correlation with evolving correlation modes coupled to the resolved flow. The modes are therefore dynamical variables rather than prescribed spatial noise shapes. Their evolution requires an additional closure, whose relation to the variational structure is made explicit below.

Starting from Luke's principle \cite{Luke67}, we introduce resolved and fluctuating harmonic potentials. The resolved component is random, since it is driven by the fluctuations. This model introduces quadratic interaction noise terms that are ill-posed at decorrelation limit. A LU balance closure needs to be introduced to obtain a well defined model at the decorrelation limit. This model can be shown to derive from a non-autonomous canonical Hamiltonian. Hence, this solution  preserves the canonical form for a prescribed fluctuating trace. It does not imply conservation of that Hamiltonian or of the total wave energy. Appendix~\ref{app-energy-balance} gives the exact energy budget for smooth finite-correlation forcing. A linearized Euler-type closure specifies the potential-mode dynamics, whereas the surface variational principle in expectation provides a formal consistency condition for the elevation modes.

The companion paper \cite{DebusscheMemin2026b} studies the kinetic implications of a linearized coupled wave--mode system. Its quantitative comparisons concern transport-induced spectral redistribution and effective interaction rates, compared with classical four-wave rates using JONSWAP spectra \cite{hasselmann1973}. They assess the relative size of these mechanisms under specified noise statistics, rather than establishing well-posedness or validating the full nonlinear system derived here. This program is reminiscent of stochastic models in wavenumber space designed to represent wave interactions more simply \cite{Apolinario-et-al-23}.

Section~2 briefly fixes the classical notation. Section~3 specifies the resolved Hamiltonian model, the mode closure and its formal variational consistency. Section~4 derives linear and shallow-water approximations. Section~5 gives the surface transport and refraction laws and a causal Ornstein--Uhlenbeck representation. \red{Section~\ref{sec:numerical-sv} assesses the stochastic Saint-Venant model numerically against a filtered Serre--Green--Naghdi reference.} The appendices recall the shape derivative of the Dirichlet-to-Neumann operator and establish the canonical formulation and energy budget.
\end{revision}

\begin{revision}
\section{Classical variational framework and notation}
\label{sec:classical}
Mainly to fix notation, we briefly recall the classical framework of Luke, Zakharov, Miles, and Craig and Sulem \cite{Craig-Sulem93,Luke67,Miles1977,Zakharov68}; see also \cite{Lannes-2013} for a comprehensive account. Here and in \S3.1, variables are dimensional. Let $\Omega_\eta=\{(x,z):x\in{\cal D}^h,\ b<z<\eta(x,t)\}$, with a flat bottom $b$, constant density $\rho$, and periodic horizontal boundaries or sufficient decay. Gradients without a subscript act on all spatial variables for bulk potentials and on $x$ for surface traces. Luke's action is
\begin{equation}
\mathcal S(\phi,\eta)=-\rho\int_{t_0}^{t_1}\int_{{\cal D}^h}\int_b^{\eta(x,t)}
\left(\partial_t\phi+\tfrac12|\nabla_{x,z}\phi|^2+gz\right)\dif z\,\dif x\,\dif t.
\label{LVF}
\end{equation}
Variations of the bulk potential give $\Delta\phi=0$, the bottom condition $\partial_z\phi|_b=0$, and, through the boundary term at the free surface, the kinematic condition. Variations of $\eta$ give Bernoulli's condition:
\begin{align}
&\left.\partial_t\eta+\nabla_x\phi\right|_\eta\cdot\nabla\eta
=\left.\partial_z\phi\right|_\eta,
\label{Kin-cond-varphi}\\
&\left.\partial_t\phi\right|_\eta+g\eta+\tfrac12\left.|\nabla_{x,z}\phi|^2\right|_\eta=0.
\label{Bernoulli-phi}
\end{align}
The notation $|_\eta$ denotes evaluation at $z=\eta(x,t)$. For the trace $\varphi=\phi\bigl(x,\eta(x,t),t\bigr)$, let
\begin{equation}
G[\eta]\varphi=\left.(\partial_z\phi-\nabla_x\phi\cdot\nabla\eta)\right|_\eta,
\qquad
B_\varphi=\frac{G[\eta]\varphi+\nabla\eta\cdot\nabla\varphi}{1+|\nabla\eta|^2}.
\label{def-DtoN}
\end{equation}
The operator $G[\eta]$ is the unnormalized Dirichlet-to-Neumann operator of the harmonic extension with the bottom condition. It is self-adjoint and non-negative on its natural domain. The chain rule gives
\begin{equation}
\left.\partial_z\phi\right|_\eta=B_\varphi,\qquad
\left.\nabla_x\phi\right|_\eta=\nabla\varphi-B_\varphi\nabla\eta,\qquad
\left.\partial_t\phi\right|_\eta=\partial_t\varphi-B_\varphi\partial_t\eta.
\label{trace-identities}
\end{equation}
In particular, \eqref{Kin-cond-varphi} reads $\partial_t\eta=G[\eta]\varphi$. After a time integration by parts and harmonic extension, Luke's action reduces, up to endpoint terms and a constant, to the Miles--Zakharov canonical action
\begin{equation}
{\mathcal S}(\varphi,\eta)=\rho \int_{t_0}^{t_1}\left[\int_{{\cal D}^h}\varphi\partial_t\eta\,\dif x-\mathcal H(\eta,\varphi)\right]\dif t,
\quad
\mathcal H=\frac12\int_{{\cal D}^h}(\varphi G[\eta]\varphi+g\eta^2)\dif x.
\label{Canonical-action}
\end{equation}
The canonical equations $\partial_t\eta=\delta_\varphi\mathcal H$ and $\partial_t\varphi=-\delta_\eta\mathcal H$ yield the kinematic condition and
\begin{equation}
\partial_t\varphi+g\eta+\frac12|\nabla\varphi|^2
-\frac12\frac{(G[\eta]\varphi+\nabla\eta\cdot\nabla\varphi)^2}{1+|\nabla\eta|^2}=0.
\label{Bernoulli-varphi}
\end{equation}
These identities are the classical ingredients needed below. The nondimensional scales are introduced in \S\ref{sec:scaling}.
\end{revision}

\section{Stochastic water waves}

\begin{revision}
\begin{newtext}
We follow the stochastic variational framework of \cite{Debussche-Memin25}
and decompose the velocity potential into resolved and fluctuating harmonic
components:
\begin{equation}
\phi=\overline\phi+\phi^\tau,\qquad
\phi^\tau(x,z,t)=\sum_{i=1}^{N}\phi_i(x,z,t)\,W_t^{\tau,i},\qquad
W_t^{\tau,i}=\int_{\mathbb R}h_\tau(t-s)\,\dif\beta_s^i .
\label{W-v}
\end{equation}
Here the $\beta^i$ are independent two-sided Brownian motions, and
$h_\tau(t)=\tau^{-1}h(t/\tau)$, where $h$ is smooth, non-negative, even,
supported on $[-1,1]$, and has unit integral.

As in \cite[eq.~(16)]{Debussche-Memin25}, the unresolved scales are assumed
to be correlated with both the past and the future of the resolved motion
over a time scale $\tau$. The temporal factors are therefore obtained by a
symmetric smoothing of white noise. Each $W^{\tau,i}_t$ is a stochastic
(Wiener) integral with a deterministic integrand: it is a stationary,
centered Gaussian process with smooth paths, which depends on the
Brownian increments over $[t-\tau,t+\tau]$. Its covariance is
\begin{equation}
\mathbb E[W_t^{\tau,i}W_t^{\tau,j}]=\Upsilon_\tau\,\delta_{ij},\qquad
\Upsilon_\tau=\frac1\tau\int_{-1}^{1}h(s)^2\,\dif s ,
\label{var}
\end{equation}
so that $W_t^{\tau,i}=\mathcal O(\tau^{-1/2})$ in root-mean-square
magnitude. Since $[\phi^\tau]=L^2T^{-1}$, the modes have units
$[\phi_i]=L^2T^{-1/2}$.

The modes $\phi_i$ are not prescribed: their evolution is coupled to the
resolved flow and to the fluctuating surface (\S\ref{sec:mode-closure}), so
that they are random and, through the noise, may depend on Brownian
increments up to time $t+\tau$. The resolved component $\overline\phi$ is
random as well. At finite $\tau$, all equations involving $\phi^\tau$ are
random partial differential equations with smooth time dependence, which
are solved pathwise; no stochastic integral is involved. Since the
regularization involves both the past and the future of the noise,
products with $W^{\tau,i}$ converge, in the decorrelation limit
$\tau\to0$, to Stratonovich products
\cite[Section~3.1 and Appendix~B]{Debussche-Memin25}; they are therefore
interpreted in the Stratonovich sense.

The Stratonovich form is, however, not convenient for computing
expectations, since the coefficients and the noise are correlated.
Following \cite[Appendix~B]{Debussche-Memin25}, we therefore use a
regularized It\^o representation, in which the noise acts through
increments over the future window $[t,t+\tau]$, independent of
$\mathcal F_t$-measurable coefficients:
\begin{equation}
\widehat W_t^{\tau,i}=\int_t^{t+\tau}\widetilde h_\tau(t-s)\,\dif\beta_s^i,
\qquad
\widetilde h_\tau(t-s)=h_\tau(t-s)+h_\tau(t+\tau-s),
\quad s\in[t,t+\tau],
\label{Ito-kernel}
\end{equation}
where $\int_t^{t+\tau}\widetilde h_\tau(t-s)\,\dif s=1$. Splitting the
symmetric window at time $t$ shows that, for a coefficient $A$ whose
noise component reads $\sum_i a_i\,\dif\beta^i$,
\begin{equation}
A(t)\,W_t^{\tau,i}=A(t)\,\widehat W_t^{\tau,i}+\tfrac12\,a_i(t)
+\text{terms vanishing as }\tau\to0 .
\label{Strato-Ito-regularized}
\end{equation}
The additional drift is the regularized counterpart of the
It\^o--Stratonovich correction; when $A$ involves the modes, it also
contains the contribution of the correlation between the modes and the
noise. The noise coefficients are unchanged by this conversion. In
particular, the fact that the temporal factors are centered does not
imply that $\phi^\tau$ is centered when the modes depend on the noise.

For clarity, we use a finite collection of modes; countable expansions
can be considered under appropriate summability and regularity
assumptions. All finite-$\tau$ equations below are understood pathwise,
for sufficiently regular solutions with positive fluid depth and no wave
breaking, and we assume
$\mathbb E\|\phi^\tau(t)\|_{L^2(\Omega_\eta)}^2<\infty$.

For finite $\tau$, the elevation has an ordinary time derivative. In the
decorrelation limit, the fluctuating velocity becomes white in time,
whereas the free surface remains a continuous semimartingale,
\begin{equation}
\dif\eta=\widetilde\eta\,\dif t+\sum_i\eta_i\circ\dif\beta_t^i ,
\end{equation}
where $\widetilde\eta$ and $\eta_i$ denote the Stratonovich drift and
noise coefficients. Assuming sufficient spatial regularity of the surface
and of the potential modes, the traces
$\varphi_i(x,t)=\phi_i(x,\eta(x,t),t)$ remain well defined, and the
fluctuating surface-potential increment is represented by
$\sum_i\varphi_i\circ\dif\beta_t^i$.
\end{newtext}

Figure~\ref{fig:variational-wave-mode-architecture} summarizes the coupled variational construction described in the following.
The pathwise principle, together with the surface-fluctuation closure  define  the resolved surface dynamics for a prescribed fluctuating trace. The principle in expectation supplies the evolution of the correlation modes through its Euler formulation, yielding a linearized Euler-type equation. The coupling acts in both directions: the resolved flow transports and deforms the correlation modes, while their reconstruction drives the free-surface evolution. This mode dynamics connects the framework with the Kraichnan--Tennekes random-sweeping picture \cite{Chen-Kraichnan-1989,Kraichnan-1964,Tennekes-1975}.
\begin{figure}
  \centering
  \includegraphics[width=\textwidth]{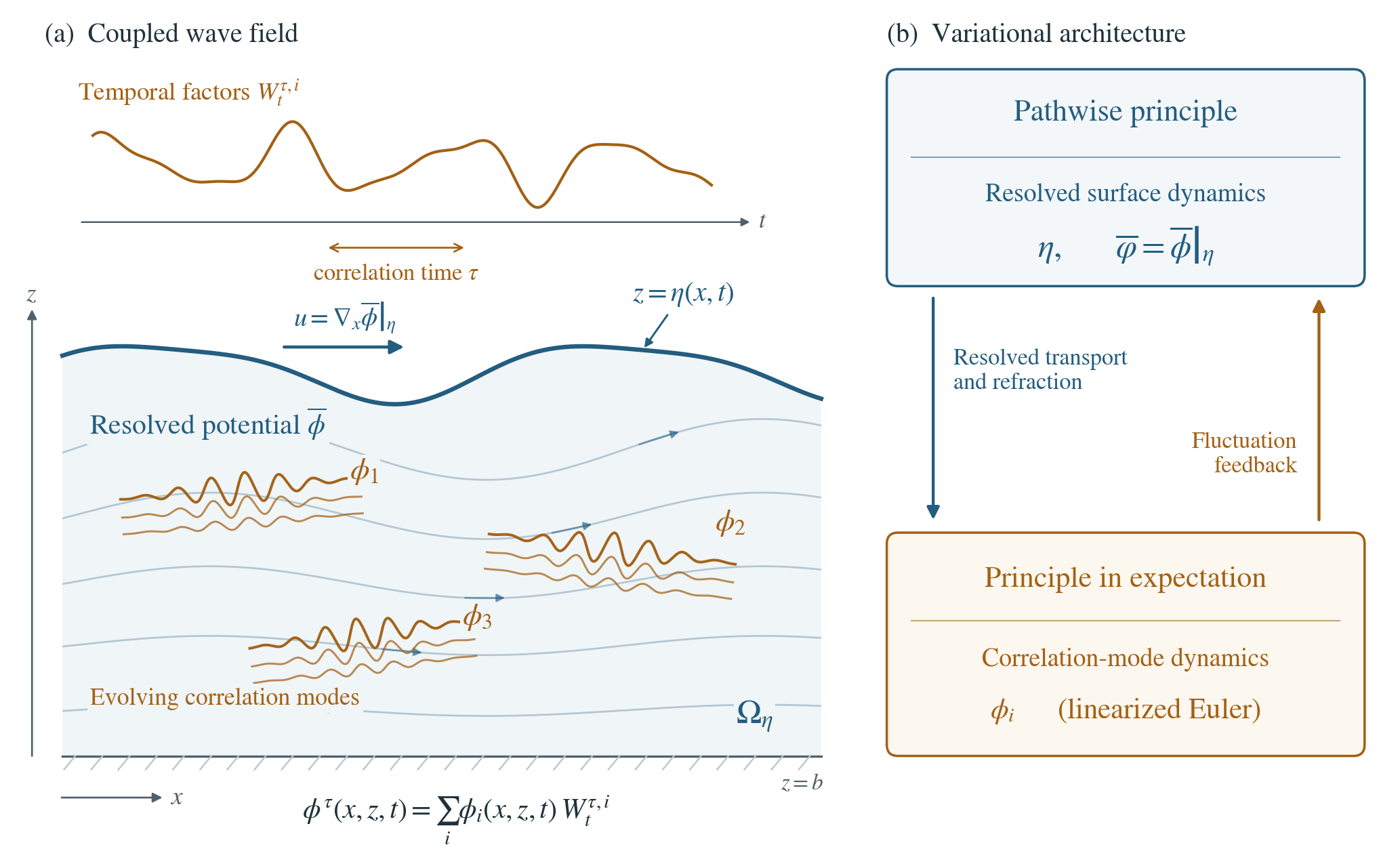}
  \caption{Schematic of the stochastic variational wave--mode framework.
  (a) The resolved free surface $\eta$ and potential $\overline\phi$
  are coupled to evolving harmonic correlation modes $\phi_i$.
  The fluctuating potential combines these spatial modes with
  regularized temporal factors $W_t^{\tau,i}$ of correlation time $\tau$.
  Blue curves indicate the resolved flow; ochre oscillatory patterns
  depict the correlation modes.
  (b) A pathwise variational principle describes the resolved surface
  dynamics, while the principle in expectation provides the
  correlation-mode dynamics through its Euler formulation.
  The resolved flow transports and refracts the modes, and the
  reconstructed fluctuations feed back on the free-surface evolution.
  The physical field and temporal trace are schematic.}
  \label{fig:variational-wave-mode-architecture}
\end{figure}

\end{revision}

\begin{revision}
\subsection{Pathwise variational principle and definition of the resolved model}
\label{sec:pathwise}
At finite correlation time, a pathwise variational principle for stochastic water waves is defined by considering  Luke's functional \eqref{LVF}, denoted again by $\mathcal S$,
\begin{equation}
\mathcal S=-\rho\int_{t_0}^{t_1}\int_{\Omega_\eta}
\left(\partial_t\phi+\tfrac12|\nabla_{x,z}\phi|^2+gz\right)\dif x\,\dif z\,\dif t,
\label{LV}
\end{equation}
motivates the construction. Both potential components are defined by harmonic extension of their traces:
\begin{align}
\Delta\overline\phi=\Delta\phi^\tau&=0\quad\hbox{in }\Omega_\eta,
\label{Laplace-Regsto}\\
\partial_z\overline\phi|_b=\partial_z\phi^\tau|_b&=0,\qquad
\overline\phi|_\eta=\overline\varphi,\quad\phi^\tau|_\eta=\varphi^\tau.
\nonumber
\end{align}
The total-potential surface conditions read
\begin{align}
&\partial_t\eta+\left.\nabla_x(\overline\phi+\phi^\tau)\right|_\eta\cdot\nabla\eta
=\left.\partial_z(\overline\phi+\phi^\tau)\right|_\eta,
\label{Selev}\\
&\partial_t\varphi+g\eta+\tfrac12|\nabla\varphi|^2
-\tfrac12\frac{(G[\eta]\varphi+\nabla\eta\cdot\nabla\varphi)^2}{1+|\nabla\eta|^2}=0,
\qquad \varphi=\overline\varphi+\varphi^\tau.
\label{Strace}
\end{align}
The velocity in \eqref{Selev} is the horizontal \emph{bulk} velocity evaluated at the surface, not the gradient of the surface trace. Equivalently, $\partial_t\eta=G[\eta](\overline\varphi+\varphi^\tau)$. It is important to remark that quadratic noise self-interactions in \eqref{Strace} obstruct a direct white-noise interpretation. It is indeed difficult to give a mathematical meaning  to those terms as they blow up at zero decorrelation time. A well posed expression at vanishing decorrelation limit  requires a closure allowing to get rid of the problematic terms. In the following we introduce a LU type closure. 

\paragraph{Fluctuation balance.}
Following the LU modeling approach \cite{memin2014fluid}, we separate
the rapidly varying fluctuation acceleration and self-interactions from
the interactions that admit a transport-noise interpretation in the
white-noise limit. This separation is used to construct a reduced model
for the resolved variables.

A pressure-free, unforced bulk balance of the form
\begin{equation}
\partial_t\phi^\tau+\frac12|\nabla_{x,z}\phi^\tau|^2=0
\label{Balance-noise}
\end{equation}
cannot be imposed together with harmonicity for general modes. Indeed,
for harmonic $\phi^\tau$,
\[
\Delta\!\left(\frac12|\nabla_{x,z}\phi^\tau|^2\right)
=|\nabla^2_{x,z}\phi^\tau|^2,
\]
so taking the Laplacian of \eqref{Balance-noise} would force
$\nabla^2_{x,z}\phi^\tau=0$. A bulk description of the fluctuation balance
must therefore retain an appropriate pressure or forcing compatible
with harmonicity. We formulate the closure directly at the free surface.

Let $s=\varphi^\tau$ and define the vertical and horizontal fluctuation
velocities evaluated at the surface by
\begin{equation}
B_s=\left.\partial_z\phi^\tau\right|_\eta
=\frac{G[\eta]s+\nabla\eta\cdot\nabla s}{1+|\nabla\eta|^2},
\qquad
U_s=\left.\nabla_x\phi^\tau\right|_\eta
=\nabla s-B_s\nabla\eta.
\label{surface-noise-velocities}
\end{equation}
Using the full kinematic condition \eqref{Selev}, the exact trace chain
rule reads
\begin{equation}
\partial_t s=\left.\partial_t\phi^\tau\right|_\eta
+B_sG[\eta]s+B_sG[\eta]\overline\varphi.
\label{surface-noise-constraint}
\end{equation}
For the unit-integral temporal regularization, the variance satisfies
$\Upsilon_\tau=\mathbb E[(W_t^{\tau,i})^2]\sim\tau^{-1}$.
At fixed spatial mode amplitudes and bounded resolved fields, the
quadratic fluctuation terms have formal size
$B_sG[\eta]s=\mathcal O(\Upsilon_\tau)$, whereas the mixed contribution
$B_sG[\eta]\overline\varphi$ is
$\mathcal O(\Upsilon_\tau^{1/2})$. The fluctuation acceleration also
contains the faster temporal derivatives
$\partial_tW_t^{\tau,i}=\mathcal O(\tau^{-3/2})$.
Although the mixed term is singular pointwise, it is linear in the
regularized noise and admits a genuine stochastic interpretation after
time integration. The quadratic fluctuation products have no direct
white-noise interpretation without an additional closure.

This ordering motivates isolating the fast fluctuation sector.
The identity $G[\eta]s=B_s-U_s\cdot\nabla\eta$ gives
\begin{equation}
\frac12\bigl(|U_s|^2+B_s^2\bigr)-B_sG[\eta]s
=\frac12|\nabla s|^2
-\frac12(1+|\nabla\eta|^2)B_s^2
=B_\eta(s),
\label{surface-noise-kinetic-identity}
\end{equation}
where
\begin{equation}
B_\eta(f)=\frac12|\nabla f|^2
-\frac12\frac{(G[\eta]f+\nabla\eta\cdot\nabla f)^2}
{1+|\nabla\eta|^2}.
\label{surface-Bernoulli-form}
\end{equation}
Consequently, the exact relation between the surface and bulk
fluctuation residuals is
\begin{equation}
\partial_t s+B_\eta(s)
=\left.\left(\partial_t\phi^\tau
+\frac12|\nabla_{x,z}\phi^\tau|^2\right)\right|_\eta
+B_sG[\eta]\overline\varphi.
\label{surface-noise-residual-identity}
\end{equation}
Motivated by scale separation at small decorrelation time,
we adopt the LU-type surface closure
\begin{equation}
\partial_t s+B_\eta(s)\approx0.
\label{surface-noise-balance}
\end{equation}
Here $\approx0$ denotes the formal balance retained when
discarding the fast fluctuation block from the resolved
Bernoulli equation. The reconstructed fluctuation remains
governed by the correlation-mode dynamics, and its residual
is given exactly by \eqref{surface-noise-residual-identity}.

\paragraph{Resolved surface equations.}
Let us substitute $\varphi=\overline\varphi+s$ into the total surface
Bernoulli equation \eqref{Strace}. Removing the fluctuation block through
\eqref{surface-noise-balance} gives
\[
\partial_t\overline\varphi+g\eta
+B_\eta(\overline\varphi+s)-B_\eta(s)=0.
\]
Expanding the quadratic form yields the resolved dynamic condition
\begin{multline}
\partial_t\overline\varphi+g\eta
+\frac12|\nabla\overline\varphi|^2
-\frac12\frac{
\bigl(G[\eta]\overline\varphi
+\nabla\eta\cdot\nabla\overline\varphi\bigr)^2}
{1+|\nabla\eta|^2}
\\
+\nabla\overline\varphi\cdot\nabla s
-\frac{
\bigl(G[\eta]\overline\varphi
+\nabla\eta\cdot\nabla\overline\varphi\bigr)
\bigl(G[\eta]s+\nabla\eta\cdot\nabla s\bigr)}
{1+|\nabla\eta|^2}=0.
\label{eq-S-varphi}
\end{multline}
The kinematic condition retains the normal velocities of both
components:
\begin{equation}
\partial_t\eta=G[\eta](\overline\varphi+s).
\label{eq-S-dyn-bound-cond}
\end{equation}
Equivalently,
\[
\partial_t\eta
+\left.\nabla_x(\overline\phi+\phi^\tau)\right|_\eta
\cdot\nabla\eta
=\left.\partial_z(\overline\phi+\phi^\tau)\right|_\eta.
\]
Equations~\eqref{eq-S-varphi}--\eqref{eq-S-dyn-bound-cond}, together
with the harmonic extensions and bottom conditions, define the
resolved model for a prescribed fluctuating trace $s$. The remaining
fluctuation contributions are linear in $s$ at fixed surface geometry,
allowing a formal transport-noise interpretation under suitable
regularity assumptions.
\paragraph{Equivalent formulation by a non-autonomous Hamiltonian.}
The resolved system
\eqref{eq-S-varphi}--\eqref{eq-S-dyn-bound-cond}
obtained above also admits a canonical Hamiltonian formulation.
For a prescribed fluctuating trace $s=s(x,t)$, let us define
\begin{equation}
\overline{\mathcal H}(\eta,\overline\varphi;s(t))
=\int_{{\cal D}^h}
\left[
\frac12\overline\varphi G[\eta]\overline\varphi
+s(t)G[\eta]\overline\varphi
+\frac12g\eta^2
\right]\dif x.
\label{resolved-H}
\end{equation}
At each time, variations with respect to $\eta$ and
$\overline\varphi$ are taken at fixed $s$. Self-adjointness of
$G[\eta]$ and its shape derivative  \cite{Lannes-2013} give
\begin{equation}
\frac{\delta\overline{\mathcal H}}{\delta\overline\varphi}
=G[\eta](\overline\varphi+s),
\qquad
\frac{\delta\overline{\mathcal H}}{\delta\eta}
=g\eta+B_\eta(\overline\varphi+s)-B_\eta(s).
\label{resolved-H-derivatives}
\end{equation}
Consequently, the canonical equations
\begin{equation}
\partial_t\eta
=\frac{\delta\overline{\mathcal H}}{\delta\overline\varphi},
\qquad
\partial_t\overline\varphi
=-\frac{\delta\overline{\mathcal H}}{\delta\eta}
\label{resolved-canonical-equations}
\end{equation}
reproduce exactly the resolved kinematic and dynamic conditions
\eqref{eq-S-dyn-bound-cond} and \eqref{eq-S-varphi}.
Thus the same reduced system can be obtained either by the surface
fluctuation closure described above or directly from the Hamiltonian
\eqref{resolved-H}.

The relation with Luke's principle is detailed in
Appendix~\ref{Reduction-Luke}. After harmonic extension and removal
of temporal endpoint terms, Luke's action per unit density differs
from the canonical action of \eqref{resolved-H} by
\[
\int_{t_0}^{t_1}
\left[
\langle s,\partial_t\eta\rangle
-\frac12\langle s,G[\eta]s\rangle
\right]\dif t.
\]
At fixed $s$, the variational derivative of these additional terms
with respect to $\eta$ is $-(\partial_t s+B_\eta(s))$.
Discarding this fluctuation block, as prescribed by the surface
closure, therefore gives the same canonical resolved equations.
When the surface balance is satisfied exactly, the corresponding
surface Euler--Lagrange equations coincide without a residual.

Since $s(t)$ is prescribed, $\overline{\mathcal H}$ depends explicitly
on time and is generally not conserved. The total energy also includes
the fluctuation self-energy. Its finite-correlation budget, together
with the role of the fluctuation-balance residual and the conditions
for pathwise conservation, is derived in
Appendix~\ref{app-energy-balance}.
The canonical structure established here concerns the resolved
variables for a prescribed fluctuating trace.

\end{revision}

\begin{revision}
\subsection{Nondimensionalization}
\label{sec:scaling}
Following \cite{Lannes-2013}, the resolved-wave scales are
\begin{equation}
\epsilon=\frac AH,\quad \mu=\frac{H^2}{L^2},\quad
\nu=\frac{\tanh(2\pi\sqrt\mu)}{2\pi\sqrt\mu},\quad
T=\frac L{\sqrt{gH\nu}},\quad
\Phi_0=\epsilon L\sqrt{\frac{gH}{\nu}}.
\label{resolved-scales}
\end{equation}
Thus $x=L\widetilde x$, $z=H\widetilde z$, $t=T\widetilde t$, $\eta=A\widetilde\eta$ and $\overline\phi=\Phi_0\widetilde{\overline\phi}$. The scaled domain is $-1<\widetilde z<\epsilon\widetilde\eta$, and the harmonic extensions satisfy $\mu\Delta_{\widetilde x}\widetilde\phi+\partial_{\widetilde z}^2\widetilde\phi=0$. The operator normalization is $G[\eta]\overline\varphi=(\Phi_0/H)G^{\mu,\epsilon}[\widetilde\eta]\widetilde{\overline\varphi}$. Tildes are omitted in nondimensional equations.
\end{revision}

We retain the resolved-wave scales introduced above: horizontal length
$L$, depth $H$, amplitude $A$ and time $T$. To scale the stochastic
contribution, we additionally introduce an amplitude scale $A_\sigma$
and a characteristic horizontal length $L_\sigma$ for the unresolved
motions. The corresponding dimensionless parameters are
\begin{equation}
\epsilon_\sigma=\frac{A_\sigma}{H},
\qquad
\mu_\sigma=\frac{H^2}{L_\sigma^2},
\qquad
r=\frac{L_\sigma}{L}.
\end{equation}
The temporal scale of the noise is specified below from gravity-wave
dispersion.

The interpretation of $A_\sigma$ depends on the physical processes represented by
the noise. When these represent an unresolved wave spectrum, $A_\sigma$
can be defined as an equivalent wave amplitude through
$E_\sigma=\tfrac12\rho gA_\sigma^2$, where $E_\sigma$ is the wave energy per unit horizontal surface area,
integrated over all unresolved wavenumbers.

\begin{revision}
Assuming that the one-dimensional wavenumber spectrum follows $E(k)=Ck^{-q}$, with $q>1$, beyond a resolved reference wavenumber $k_p$, we approximate the resolved band energy by $E_p\simeq E(k_p)\Delta k_p$, with $\Delta k_p=\beta k_p$. Integrating the unresolved tail beyond $k_c\geq k_p$ gives
\begin{equation}
\frac{A_\sigma}{A}=\left(\frac{E_\sigma}{E_p}\right)^{1/2}
\simeq\left[\frac{1}{\beta(q-1)}\left(\frac{k_c}{k_p}\right)^{1-q}\right]^{1/2},
\qquad E_p=\tfrac12\rho gA^2.
\label{spectral-amplitude}
\end{equation}
For $k_c\geq k_p$, the estimate is bounded above by $[\beta(q-1)]^{-1/2}$. Thus a steep tail with an order-one reference bandwidth gives $A_\sigma\lesssim A$; it does not imply an arbitrarily large unresolved amplitude. Regimes with $A_\sigma\gtrsim A$ describe energetic unresolved processes not accounted for by this tail estimate, such as additional storm-generated waves; they do not follow from \eqref{spectral-amplitude}. The exponent $q$ is distinct from the memory parameter $\m$ below. We refer to $\epsilon_\sigma=A_\sigma/H$ as the stochastic amplitude parameter.

The surface is described by the single elevation field $\eta=A\widetilde\eta$; $A_\sigma$ sets the scale of the fluctuating potential. For an unresolved wavelength $L_\sigma$, linear gravity-wave dispersion gives
\begin{equation}
\nu_\sigma=\frac{\tanh(2\pi\sqrt{\mu_\sigma})}{2\pi\sqrt{\mu_\sigma}},
\qquad T_\sigma=\frac{L_\sigma}{\sqrt{gH\nu_\sigma}},
\qquad\mu_\sigma=\frac{H^2}{L_\sigma^2}.
\end{equation}
Here $T_\sigma=2\pi/\omega_\sigma$ for $k_\sigma=2\pi/L_\sigma$. The unresolved waves may be in deep water even when the resolved waves are shallow.
\end{revision}

We choose the correlation time of the regularized noise to equal this
unresolved-wave period, $\tau=T_\sigma$. This is a modeling choice
that links temporal persistence to wavelength. Relative to the
resolved-wave time $T=L/\sqrt{gH\nu}$, it gives the dimensionless memory
parameter
\begin{equation}
\m=\frac{\tau}{T}=\frac{T_\sigma}{T}
=r\sqrt{\frac{\nu}{\nu_\sigma}}.
\end{equation}
Small values $\m\ll1$ describe short-memory forcing, whereas
$\m=\mathcal O(1)$ describes forcing that remains correlated over a
resolved-wave time scale. For a common depth, $L_\sigma\leq L$ implies
$\m\leq1$. When both components are in shallow water, $\m\simeq r$;
when both are in deep water, $\m\simeq\sqrt r$. The role of this temporal
persistence in energy transfer is examined in the companion kinetic
paper.

We use the same linear-wave potential scaling as for the resolved
component, evaluated at the unresolved amplitude and wavelength:
\begin{equation}
\phi^\tau=\Phi_0^\sigma\widetilde\phi^\tau,
\qquad
\Phi_0^\sigma
=\epsilon_\sigma L_\sigma\sqrt{\frac{gH}{\nu_\sigma}}
=gA_\sigma T_\sigma.
\end{equation}
The surface trace $\varphi^\tau$ is scaled by the same factor. Following
the noise definition \eqref{W-v}, let us write
$\phi^\tau=\sum_j\phi_jW_t^{\tau,j}$. Since
$W_t^\tau=\mathcal O(T_\sigma^{-1/2})$, the spatial correlation modes
require the scaling
\begin{equation}
\phi_j=\Phi_{j,0}^\sigma\widetilde\phi_j,
\qquad
\Phi_{j,0}^\sigma=\Phi_0^\sigma\sqrt{T_\sigma}.
\end{equation}
The additional factor $\sqrt{T_\sigma}$ accounts for the temporal
normalization of the regularized noise, consistently with
$[\phi^\tau]=L^2T^{-1}$ and $[\phi_j]=L^2T^{-1/2}$.

For the nondimensional equations, it is convenient to collect the
amplitude, length and depth factors into
\begin{equation}
\epsilon_\sigma^r=\epsilon_\sigma r,
\qquad
\gamma_\sigma=\frac{\epsilon_\sigma}{\sqrt{\nu_\sigma}},
\qquad
\gamma_\sigma^r=r\gamma_\sigma
=\frac{\epsilon_\sigma}{\sqrt\nu}\m.
\end{equation}
In particular, $\Phi_0^\sigma=L\sqrt{gH}\,\gamma_\sigma^r$,
where $\gamma_\sigma^r$ measures the fluctuating-potential amplitude
relative to the reference potential scale $L\sqrt{gH}$, formed from
the resolved horizontal length $L$ and the velocity scale $\sqrt{gH}$.
 Its expression in terms of $\m$ makes explicit the relation between amplitude and
temporal scale under the adopted prescription $\tau=T_\sigma$.

Both potentials are expressed in the common coordinates
$\mathbf x=L\widetilde{\mathbf x}$ and $z=H\widetilde z$. Consequently,
they share the same dimensionless Dirichlet-to-Neumann operator,
determined by the common free-surface geometry:
\begin{equation}
G[\eta]\varphi^\tau
=\frac{\Phi_0^\sigma}{H}
G^{\mu,\epsilon}[\widetilde\eta]\widetilde\varphi^{\,\tau}.
\end{equation}
Dropping tildes on dimensionless quantities, this operator reads
\begin{equation}
G^{\mu,\epsilon}[\eta]\varphi^\tau
=\left(
\partial_z\phi^\tau
-\mu\epsilon\nabla_x\phi^\tau\cdot\nabla_x\eta
\right)_{z=\epsilon\eta}.
\end{equation}
The shorter unresolved scale is represented by the spatial variation
of the fluctuating potential and by the scaling coefficients above.
The resulting nondimensional free-surface equations, together with
the correlation-mode equation stated below, are gathered in
\S\ref{sec:final-models}.
\paragraph{Remark on rotational noise.} The large-scale flow may remain potential while the prescribed fluctuating velocity is rotational \cite{Street-2023}. In that setting the interface transport contains both the horizontal and vertical noise velocities. Here we retain potential noise: its bulk harmonic extension supplies the surface covariance required for the mode equations below.

\begin{revision}
\subsection{Mode closure and variational consistency in expectation}
\label{sec:mode-closure}
\paragraph{A linearized Euler-type closure.}
The multiscale construction of \cite{Debussche-Memin25} couples a
pathwise variational principle for the full flow with a variational
principle in expectation for the noise correlation modes. For the
incompressible Euler equations, this construction leads to stochastic
resolved dynamics coupled to a linearized Euler-type equation for the
noise modes. Here we adopt the potential-flow form of this modal
closure. In dimensional variables, let us denote
$\overline U=\nabla_{x,z}\overline\phi$ and
$U_i=\nabla_{x,z}\phi_i$. The mode equations read
\begin{align}
\partial_tU_i
+(\overline U\cdot\nabla)U_i
+(U_i\cdot\nabla)\overline U
&=-\nabla\pi_i,
\qquad \nabla\cdot U_i=0,
\label{linearized-mode-Euler}\\
\partial_t\phi_i
+\nabla\overline\phi\cdot\nabla\phi_i
+\pi_i&=0.
\label{mode-Bernoulli}
\end{align}
The second equation follows by spatial integration, with the
time-dependent integration constant absorbed into $\pi_i$.
Here $\pi_i$ is the modal pressure enforcing harmonicity of $\phi_i$.
For smooth harmonic potentials, taking the divergence of
\eqref{linearized-mode-Euler} gives
\begin{equation}
\Delta\pi_i
=-2\nabla^2\overline\phi:\nabla^2\phi_i,
\qquad
\pi_i|_\eta=0,
\qquad
\partial_z\pi_i|_b=0.
\label{pressure-mart}
\end{equation}
Assuming constant atmospheric pressure, we impose pressure equilibrium
mode by mode at the instantaneous free surface, giving
$\pi_i|_\eta=0$. The bottom condition follows from the impermeability
conditions for the resolved and modal velocities at the flat bottom.
These boundary conditions complete the bulk modal closure. The surface
variational principle below supplies the corresponding relation
between the potential modes and the elevation modes.

Let us define the resolved vertical velocity $\overline B$, resolved
horizontal velocity $u$, and modal vertical velocity $B_i$ at the
free surface by
\begin{equation}
\overline B
=\frac{G[\eta]\overline\varphi
+\nabla\eta\cdot\nabla\overline\varphi}{1+|\nabla\eta|^2},
\quad
u=\nabla\overline\varphi-\overline B\nabla\eta,
\quad
B_i=\frac{G[\eta]\varphi_i
+\nabla\eta\cdot\nabla\varphi_i}{1+|\nabla\eta|^2}.
\label{surface-mode-velocities}
\end{equation}
The moving-trace identities are
\begin{equation}
\partial_t\varphi_i
=\left.\partial_t\phi_i\right|_\eta+B_i\partial_t\eta,
\qquad
\left.\nabla\overline\phi\cdot\nabla\phi_i\right|_\eta
=u\cdot\nabla\varphi_i+B_iG[\eta]\overline\varphi.
\label{mode-trace-chain}
\end{equation}
Using $\pi_i|_\eta=0$ and the kinematic condition
\eqref{eq-S-dyn-bound-cond} in \eqref{mode-Bernoulli}, together with
\eqref{mode-trace-chain}, therefore yields
\begin{equation}
\partial_t\varphi_i
+u\cdot\nabla\varphi_i
-B_iG[\eta]\varphi^\tau=0.
\label{varphi-i-sto}
\end{equation}
This equation is linear in each mode when
$(\eta,\overline\varphi,\varphi^\tau)$ are prescribed. The reconstructed
coupled family is nonlinear because $\varphi^\tau$ itself depends on
the evolving modes. The reconstructed fluctuation is not required to
satisfy the separate balance \eqref{surface-noise-balance}.
The associated energy exchange is quantified by
\eqref{app-mode-energy-residual} in
Appendix~\ref{app-energy-balance}.

\paragraph{Variational principle in expectation for the correlation modes.}
\begin{newtext}
To evaluate the expectations entering the mode calculation, we use the
regularized It\^o representation \eqref{Ito-kernel}--\eqref{Strato-Ito-regularized}.
As in \cite[Section~3.1]{Debussche-Memin25}, the coefficients are frozen
at time $t$ over the window $[t,t+\tau]$; the resulting error vanishes as
$\tau\to0$. The coefficients are then $\mathcal F_t$-measurable, whereas
$\widehat W_t^{\tau,i}$ only involves the Brownian increments over
$[t,t+\tau]$. Independence of these increments gives
\begin{equation}
\mathbb E[\widehat W_t^{\tau,i}\mid\mathcal F_t]=0,
\qquad
\mathbb E[\widehat W_t^{\tau,i}\widehat W_t^{\tau,j}\mid\mathcal F_t]
=\widehat\Upsilon_\tau\,\delta_{ij},
\qquad
\widehat\Upsilon_\tau=\int_t^{t+\tau}\widetilde h_\tau(t-s)^2\,\dif s
=\mathcal O(\tau^{-1}).
\label{Ito-conditional-moments}
\end{equation}
Consequently, for adapted coefficients with the required integrability,
\[
\mathbb E[A_i(t)\widehat W_t^{\tau,i}]=0,
\qquad
\mathbb E[A_{ij}(t)\widehat W_t^{\tau,i}\widehat W_t^{\tau,j}]
=\widehat\Upsilon_\tau\,\delta_{ij}\,\mathbb E[A_{ij}(t)].
\]
These identities follow from conditional expectation and do not require
the coefficients to be deterministic or independent of the past
evolution.

In this representation, the regularized elevation rate and potential
fluctuation read
\begin{equation}
\partial_t\eta=\widetilde\eta^{\mathrm I}
+\sum_i\eta_i\widehat W_t^{\tau,i},
\qquad
\varphi^{\tau,\mathrm I}=\sum_i\varphi_i\widehat W_t^{\tau,i}.
\label{elevation-rate-modes}
\end{equation}
By \eqref{Strato-Ito-regularized}, the drift $\widetilde\eta^{\mathrm I}$
differs from its Stratonovich counterpart by a finite-variation
correction, while the noise coefficients $\eta_i$ are the same in both
representations. In the semimartingale limit,
\begin{equation}
\dif\bigl\langle\eta,\beta^i\bigr\rangle_t=\eta_i\,\dif t .
\label{elevation-mode-covariation}
\end{equation}
The correction does not contribute to this covariation and therefore does
not affect the identification of the elevation noise coefficients.
\end{newtext}

Consider the expected total surface action
\begin{equation}
\mathbb S
=\mathbb E\int_{t_0}^{t_1}\int_{{\cal D}^h}
\left(
\varphi\partial_t\eta
-\frac12\varphi G[\eta]\varphi
-\frac12g\eta^2
\right)\dif x\,\dif t.
\label{Exp-ZCS-functional}
\end{equation}
Let us vary a potential mode, treating $\eta$ as an independent
variable of the action. In the It\^o representation,
$\delta\varphi=\delta\varphi_i\widehat W_t^{\tau,i}$, where
$\delta\varphi_i$ is adapted. Self-adjointness of $G[\eta]$ gives
\[
\delta_{\varphi_i}\mathbb S
=\mathbb E\int_{t_0}^{t_1}
\left\langle
\delta\varphi_i\widehat W_t^{\tau,i},
\partial_t\eta-G[\eta]\varphi
\right\rangle\dif t.
\]
The residual decomposes as
\[
\partial_t\eta-G[\eta]\varphi
=\widetilde\eta^{\mathrm I}-G[\eta]\overline\varphi
+\sum_j\bigl(\eta_j-G[\eta]\varphi_j\bigr)
\widehat W_t^{\tau,j}.
\]
The contribution of the first two terms to the variation has zero
expectation by \eqref{Ito-conditional-moments}. Thus the terms linear
in the centered It\^o noise vanish.
Equivalently, the expected action splits into
$\mathbb S=\mathbb S_0+\mathbb S_\varphi^\tau$, where
$\mathbb S_0$ is independent of the potential modes in this variation
and
\begin{equation}
\mathbb S_\varphi^\tau
=\widehat\Upsilon_\tau\,
\mathbb E\int_{t_0}^{t_1}\int_{{\cal D}^h}
\sum_i\left(
\varphi_i\eta_i
-\frac12\varphi_iG[\eta]\varphi_i
\right)\dif x\,\dif t.
\label{Exp-ZCS-functional-2}
\end{equation}
The two functionals therefore have the same first variation with
respect to the potential modes. Applying the conditional covariance
identity yields
\begin{equation}
\delta_{\varphi_i}\mathbb S
=\delta_{\varphi_i}\mathbb S_\varphi^\tau
=\widehat\Upsilon_\tau\,
\mathbb E\int_{t_0}^{t_1}
\langle\delta\varphi_i,
\eta_i-G[\eta]\varphi_i\rangle\dif t,
\label{expected-action-mode-variation}
\end{equation}
where $\langle f,k\rangle=\int_{{\cal D}^h}fk\,\dif x$.
Since $\widehat\Upsilon_\tau>0$, stationarity of $\mathbb S$ with
respect to arbitrary admissible adapted mode variations yields
\begin{equation}
\eta_i=G[\eta]\varphi_i.
\label{eta-i-constraint}
\end{equation}
 This relation agrees with the noise coefficient in the
kinematic equation and holds in both Stratonovich and It\^o
representations.  

The evolution of the potential modes is governed by
\eqref{mode-Bernoulli}, whose surface form is \eqref{varphi-i-sto}.
For finite correlation time, the nondimensional form of
\eqref{varphi-i-sto} reads
\begin{multline}
\partial_t\varphi_i
+\frac\epsilon\nu\nabla\overline\varphi\cdot\nabla\varphi_i
-\frac{\epsilon^2\mu}{\nu}
\frac{(\nabla\varphi_i\cdot\nabla\eta)
(\mu^{-1}G^{\mu,\epsilon}[\eta]\overline\varphi
+\epsilon\nabla\overline\varphi\cdot\nabla\eta)}
{1+\epsilon^2\mu|\nabla\eta|^2}
\\
-\frac{\epsilon_\sigma\m\mu}{\nu}
\frac{(\mu^{-1}G^{\mu,\epsilon}[\eta]\varphi_i
+\epsilon\nabla\varphi_i\cdot\nabla\eta)
\mu^{-1}G^{\mu,\epsilon}[\eta]\varphi^\tau}
{1+\epsilon^2\mu|\nabla\eta|^2}=0.
\label{varphi-i-sto-adim}
\end{multline}
The last term couples the vertical mode velocity to the fluctuating
normal surface velocity. It is also the source of a potential loss
of regularity discussed below.

\end{revision}

\begin{revision}
\subsection{Final stochastic water wave models}\label{sec:final-models}

The finite-correlation model consists of the resolved Hamiltonian equations coupled to the mode closure. The full total-potential equations are recalled first as a reference.
\paragraph{Finite-correlation system}
\begin{itemize}
\item \textbf{Unclosed total-potential equations (reference).}
The total potential satisfies the dimensionless harmonic-extension problem

\begin{align}
\mu\Delta_x\phi+\partial_z^2\phi
&=0\text{ in }\Omega_t,
\label{SLaplace-fin1}\\
\partial_z\phi
&=0
\text{ at }z=-1,
\nonumber\\
\phi\bigl(x,\epsilon\eta(x,t),t\bigr)
&=\varphi(x,t)
\text{ at }z=\epsilon\eta(x,t).
\nonumber
\end{align}

The unclosed kinematic and dynamic surface conditions are
\begin{equation}
\partial_t\eta=\frac1{\mu\nu}G^{\mu,\epsilon}[\eta]\varphi,
\label{eq-surf-adim}
\end{equation}
\begin{equation}
\partial_t\varphi+\eta+\frac\epsilon{2\nu}|\nabla\varphi|^2
-\frac{\epsilon\mu}{2\nu}
\frac{(\mu^{-1}G^{\mu,\epsilon}[\eta]\varphi+\epsilon\nabla\eta\cdot\nabla\varphi)^2}
{1+\epsilon^2\mu|\nabla\eta|^2}=0.
\end{equation}
Here $\varphi$ is the total trace scaled by $\Phi_0$. With the separate scaling of \S\ref{sec:scaling}, it equals $\overline\varphi+(\epsilon_\sigma\m/\epsilon)\varphi^\tau$. These reference equations are not generally satisfied by the closed system below.

\item \textbf{Resolved Hamiltonian model for a prescribed fluctuating trace.}
The resolved and fluctuating potentials have separate harmonic extensions with the common geometry. The nondimensional kinematic equation is

\begin{equation}
\partial_t\eta
=
\frac{1}{\mu\nu}
G^{\mu,\epsilon}[\eta]\overline{\varphi}
+
\frac{\epsilon_\sigma\m}
{\epsilon\mu\nu}
G^{\mu,\epsilon}[\eta]\varphi^\tau,
\label{eq-S-eta-final-adim}
\end{equation}
while the large-scale dynamic boundary condition reads
\begin{multline}
\partial_t\overline{\varphi}
+\eta
+\frac{\epsilon}{2\nu}
|\nabla\overline{\varphi}|^2
-
\frac{\epsilon\mu}{2\nu}
\frac{
\left(
\frac{1}{\mu}
G^{\mu,\epsilon}[\eta]\overline{\varphi}
+
\epsilon\nabla\eta
\boldsymbol{\bcdot}
\nabla\overline{\varphi}
\right)^2
}{
1+\epsilon^2\mu|\nabla\eta|^2
}
\\
+
\frac{\epsilon_\sigma\m}{\nu}
\nabla\overline{\varphi}
\boldsymbol{\bcdot}
\nabla\varphi^\tau
-
\frac{\epsilon_\sigma\m\mu}{\nu}
\frac{
\left(
\frac{1}{\mu}
G^{\mu,\epsilon}[\eta]\overline{\varphi}
+
\epsilon\nabla\eta
\boldsymbol{\bcdot}
\nabla\overline{\varphi}
\right)
\left(
\frac{1}{\mu}
G^{\mu,\epsilon}[\eta]\varphi^\tau
+
\epsilon\nabla\eta
\boldsymbol{\bcdot}
\nabla\varphi^\tau
\right)
}{
1+\epsilon^2\mu|\nabla\eta|^2
}
=0.
\label{eq-S-varphi-final-adim}
\end{multline}

\item \textbf{Small-scale component (given large-scale dynamics).} \\
The noise potential modes satisfy the dimensionless Laplace problem
\begin{align}
\mu\Delta_x\phi_i+\partial_z^2\phi_i
&=0
\text{ in }\Omega_t,
\label{SLaplace-noise-fin1}\\
\partial_z\phi_i
&=0
\text{ at }z=-1,
\nonumber\\
\phi_i\bigl(x,\epsilon\eta(x,t),t\bigr)
&=\varphi_i(x,t)\text{ at }z=\epsilon\eta(x,t).
\nonumber
\end{align}
Constraint on the surface elevation correlation functions:
\begin{equation}
\eta_i    =  G^{\mu,\epsilon}[\eta] \varphi_i,
\end{equation}
where any dimensional prefactors have been absorbed into the definition of \(\eta_i\).

The mode closure in nondimensional surface variables is

\begin{multline}
\partial_t\varphi_i
+
\frac{\epsilon}{\nu}
\nabla\overline{\varphi}
\boldsymbol{\bcdot}
\nabla\varphi_i
-
\frac{\epsilon^2\mu}{\nu}
\frac{
\left(
\nabla\varphi_i
\boldsymbol{\bcdot}
\nabla\eta
\right)
\left(
\frac{1}{\mu}
G^{\mu,\epsilon}[\eta]\overline{\varphi}
+
\epsilon\nabla\overline{\varphi}
\boldsymbol{\bcdot}
\nabla\eta
\right)
}{
1+\epsilon^2\mu|\nabla\eta|^2
}
\\
-
\frac{\epsilon_\sigma\m\mu}{\nu}
\frac{
\left(
\frac{1}{\mu}
G^{\mu,\epsilon}[\eta]\varphi_i
+
\epsilon\nabla\varphi_i
\boldsymbol{\bcdot}
\nabla\eta
\right)
\frac{1}{\mu}
G^{\mu,\epsilon}[\eta]\varphi^\tau
}{
1+\epsilon^2\mu|\nabla\eta|^2
}
=0.
\label{varphi-i-final-adim}
\end{multline}
\end{itemize}

\paragraph{Time-decorrelated system}
We now consider the corresponding white-in-time model.
The Stratonovich interpretation is consistent with Wong--Zakai-type
convergence results for temporally regularized transport noise
\cite{Debussche-Pappalattera2023,
Flandoli-Pappalettera-2023}.
The correlation modes retain their dynamical evolution.
Returning to dimensional variables, the fluctuating potential
increment is written
\begin{equation}
\dif\phi^\sigma
=
\phi_i\circ\dif\beta_t^i.
\label{potential-increment-white}
\end{equation}
\begin{itemize}
\item \textbf{Large-scale component}\\
The large-scale potential satisfies the Laplace potential flow condition
\begin{align}
\Delta\overline{\phi}
&=0\text{ in }\Omega_t,
\label{SLaplace-fin2}\\
\partial_z\overline{\phi}
&=0\text{ at }z=b,
\nonumber\\
\overline{\phi}\bigl(x,\eta(x,t),t\bigr)
&=\overline{\varphi}(x,t)
\text{ at }z=\eta(x,t).
\nonumber
\end{align}
The kinematic boundary condition is
\begin{equation}
\dif\eta
=
G[\eta]\overline{\varphi}\,\dif t
+
G[\eta]\varphi_i\circ\dif\beta_t^i.
\label{Selevfin2}
\end{equation}

The large-scale dynamic boundary condition is
\begin{multline}
\dif\overline{\varphi}
+
\left[
g\eta
+
\frac{1}{2}|\nabla\overline{\varphi}|^2
-
\frac{1}{2}
\frac{
\left(
G[\eta]\overline{\varphi}
+
\nabla\eta
\boldsymbol{\bcdot}
\nabla\overline{\varphi}
\right)^2
}{
1+|\nabla\eta|^2
}
\right]\dif t
\\
+
\left[
\nabla\overline{\varphi}
\boldsymbol{\bcdot}
\nabla\varphi_i
-
\frac{
\left(
G[\eta]\overline{\varphi}
+
\nabla\eta
\boldsymbol{\bcdot}
\nabla\overline{\varphi}
\right)
\left(
G[\eta]\varphi_i
+
\nabla\eta
\boldsymbol{\bcdot}
\nabla\varphi_i
\right)
}{
1+|\nabla\eta|^2
}
\right]
\circ\dif\beta_t^i
=0.
\label{eq-S-varphi-sto}
\end{multline}

\item \textbf{Dynamics of the noise correlation modes for a given large-scale flow.}

The noise potential modes satisfy
\begin{align}
\Delta\phi_i
&=0\text{ in }\Omega_t,
\label{SLaplace-noise-fin2}\\
\partial_z\phi_i
&=0
\text{ at }z=b,
\nonumber\\
\phi_i\bigl(x,\eta(x,t),t\bigr)
&=\varphi_i(x,t)
\text{ at }z=\eta(x,t).
\nonumber
\end{align}

The coefficients of the Stratonovich surface-elevation noise satisfy

\begin{equation}
\eta_i  = G[\eta] \varphi_i,
\end{equation}

The formal surface equation for the noise potential modes is

\begin{multline}
\dif\varphi_i
+
\left[
\nabla\overline{\varphi}
\boldsymbol{\bcdot}
\nabla\varphi_i
-
\frac{
\left(
\nabla\varphi_i
\boldsymbol{\bcdot}
\nabla\eta
\right)
\left(
G[\eta]\overline{\varphi}
+
\nabla\overline{\varphi}
\boldsymbol{\bcdot}
\nabla\eta
\right)
}{
1+|\nabla\eta|^2
}
\right]\dif t
\\
-
\frac{
G[\eta]\varphi_i
+
\nabla\varphi_i
\boldsymbol{\bcdot}
\nabla\eta
}{
1+|\nabla\eta|^2
}
\,
\eta_j\circ\dif\beta_t^j
=0.
\label{varphi-i-white}
\end{multline}

\end{itemize}

\paragraph{Regularity limitation.}
Equation~\eqref{varphi-i-white} contains the coupling
$B_i\eta_j\circ\dif\beta_t^j$. For a flat surface at $z=0$
and infinite depth, the decaying harmonic extension satisfies,
for $k\neq0$,
\[
\widehat{\phi_i}(k,z,t)
=e^{|k|z}\widehat{\varphi_i}(k,t),
\qquad z<0,
\]
so that $B_i=|D|\varphi_i$. Decay at depth is therefore already
incorporated in this surface operator. Locally freezing the surface
coefficient $\eta_j=c$ and suppressing the other terms gives
\[
\dif\varphi_i=c|D|\varphi_i\circ\dif\beta_t^j,
\qquad
\widehat{\varphi_i}(k,t)
=\exp\!\left(c|k|(\beta_t^j-\beta_0^j)\right)
\widehat{\varphi_i}(k,0).
\]
Whenever $c(\beta_t^j-\beta_0^j)>0$, this multiplier grows
exponentially with $|k|$ and is unbounded on general Sobolev data.
Here the constant $c$ represents a local freezing of the boundary
coefficient.
This diagnostic reflects upward harmonic continuation as the surface 
moves. The decay as $z\to-\infty$ alone does not remove that mechanism.
A well-posedness analysis of the coupled system should therefore
account for the evolving surface geometry and determine conditions
on the spatial regularity or spectrum of the noise modes that
control this high-frequency amplification.
The corresponding normal-velocity coupling is already present at
finite $\tau$ in $B_iG[\eta]\varphi^\tau$.
We leave the well-posedness of \eqref{varphi-i-white} and the
complete coupled It\^o conversion open; the latter must include
both the moving geometry and the stochastic mode dependence.
\end{revision}

\begin{revision}
\section{Reduced systems}
\label{sec:reduced}
We first give the linear Airy model at finite depth, and then derive nonlinear shallow-water models from \eqref{eq-S-eta-final-adim}, \eqref{eq-S-varphi-final-adim} and \eqref{varphi-i-final-adim}. These constitute only formal truncations for smooth solutions. All nonlinear shallow-water expansions require $\mu\ll1$ and adequate control of the spatial derivatives of the noise. If its wavelength is $L_\sigma=rL$, a joint long-wave expansion also requires $\mu_\sigma=\mu/r^2\ll1$. One cannot apply a uniform $\mu$-expansion to arbitrarily short noise modes.

For a dimensionless trace $f$, let $\phi_f$ be its harmonic extension and define
\begin{equation}
{\cal V}_\eta f=\frac1h\int_{-1}^{\epsilon\eta}\nabla_x\phi_f\,\dif z,
\quad h=1+\epsilon\eta,
\quad G^{\mu,\epsilon}[\eta]f=-\mu\nabla\cdot(h{\cal V}_\eta f).
\label{depth-average}
\end{equation}
Let $\overline V={\cal V}_\eta\overline\varphi$, $V_i={\cal V}_\eta\varphi_i$ and $V^\tau={\cal V}_\eta\varphi^\tau$. The latter includes the nondimensional temporal factors in the noise reconstruction. Since the two potentials use different amplitude scales, the total velocity in resolved units is $\overline V+\alpha V^\tau$, where $\alpha=\epsilon_\sigma\m/\epsilon$. The classical long-wave expansion \cite{Lannes-2013} is
\begin{equation}
{\cal V}_\eta f=\nabla f-\mu\mathbb T[h]\nabla f+\mathcal O(\mu^2),
\qquad \mathbb T[h]V=-\frac1{3h}\nabla(h^3\nabla\cdot V).
\label{depth-average-expansion}
\end{equation}
This is an asymptotic expansion under regularity assumptions, not a claim of convergence of an infinite series.

\subsection{Linear and long-wave approximations}
\subsubsection{Airy waves: small amplitude at finite depth}
For $\epsilon\ll1$, $\epsilon_\sigma\m=\mathcal O(\epsilon)$ and bounded appropriately scaled spatial derivatives, linearization at $\eta=0$ gives
\begin{equation}
\partial_t\overline\varphi=-\eta,\qquad
\partial_t\eta={\cal D}_\mu\overline\varphi+\alpha{\cal D}_\mu\varphi^\tau,
\qquad \partial_t\varphi_i=0,
\quad {\cal D}_\mu=\frac{G^{\mu,0}}{\mu\nu}.
\label{eq-adim-Airy-linear}
\end{equation}
The bulk extensions satisfy $\mu\Delta_x\phi+\partial_z^2\phi=0$ on $-1<z<0$ with a homogeneous bottom Neumann condition. Let us note that the identity $G^{\mu,0}f=\partial_z\phi_f|_0$ is exact at a flat surface. It is not a shallow-water approximation. Its Fourier symbol and the dimensionless Airy frequency are
\begin{equation}
G^{\mu,0}_k=\sqrt\mu|k|\tanh(\sqrt\mu|k|),\qquad
D_\mu(k)=\frac{|k|\tanh(\sqrt\mu|k|)}{\sqrt\mu\nu},
\qquad\omega_k^2=D_\mu(k).
\label{Airy-symbol}
\end{equation}
In dimensional units, this corresponds to $\omega_{\rm dim}^2=g|k_{\rm dim}|\tanh(H|k_{\rm dim}|)$. In the genuine shallow-water limit $\sqrt\mu|k|\ll1$, $G^{\mu,0}_k\simeq\mu|k|^2$ and $\omega_{\rm dim}^2\simeq gH|k_{\rm dim}|^2$.

If the resolved and unresolved waves are both shallow, $\m\simeq r$, and $\epsilon_\sigma\sim\epsilon$, then $\alpha\simeq r$. Thus the noise enters through an $\mathcal O(r)$ linear forcing of the kinematic condition, while the modes decouple at leading order. The condition $r<1$ alone does not justify omitting this forcing.

For a periodic horizontal domain and $k\ne0$, define
\begin{equation}
a_k=\frac12\left(\widehat\eta_k+\frac{iD_\mu(k)}{\omega_k}\widehat{\overline\varphi}_k\right).
\end{equation}
Then the complete forced normal-mode solution is
\begin{align}
a_k(t)&=e^{-i\omega_kt}a_k(0)
+\frac\alpha2D_\mu(k)\int_0^t e^{-i\omega_k(t-s)}\widehat{\varphi^\tau}_k(s)\dif s,
\label{Airy-Duhamel}\\
\widehat\eta_k&=a_k+a_{-k}^*,\qquad
\widehat{\overline\varphi}_k=-\frac{i\omega_k}{D_\mu(k)}(a_k-a_{-k}^*).
\nonumber
\end{align}
The zero Fourier mode is obtained directly from \eqref{eq-adim-Airy-linear}. The harmonic reconstructions use the common dimensionless coordinates:
\begin{equation}
\overline\phi(x,z,t)=\sum_k\widehat{\overline\varphi}_k(t)
\frac{\cosh(\sqrt\mu|k|(z+1))}{\cosh(\sqrt\mu|k|)}e^{ik\cdot x},
\quad
\phi_j(x,z)=\sum_k\widehat{\varphi_j}_k
\frac{\cosh(\sqrt\mu|k|(z+1))}{\cosh(\sqrt\mu|k|)}e^{ik\cdot x}.
\end{equation}
The noise-mode label $j$ and Fourier index $k$ are distinct. At this order, stochasticity enters through an additive
random forcing of the kinematic free-surface condition. The coupling between surface elevation and resolved potential
then produces a stochastic wave response. Within the shallow-water long-wave scaling, this provides a
minimal stochastic extension of the classical deterministic
Airy-wave model for very long waves.

For comparison, if $L_\sigma\ll H\ll L$, the scales of \S\ref{sec:scaling} give $\nu\simeq1$, $\nu_\sigma\simeq L_\sigma/(2\pi H)$, and hence
\[
\m\simeq\frac{\sqrt{2\pi HL_\sigma}}{L},\qquad
\epsilon_\sigma\m\simeq\frac{A_\sigma}{L}\sqrt{\frac{2\pi L_\sigma}{H}}.
\]
Making the stochastic-advection coefficient at the resolved scale
order one would require
$A_\sigma\sim L\sqrt{H/(2\pi L_\sigma)}$.
As $L_\sigma/H\to0$, this implies $A_\sigma\gg L\gg H$,
which is incompatible with the assumed small-amplitude regime.

Consequently, within the shallow-water long-wave scaling
considered here, the Airy model retains its classical linear
structure. Unresolved fluctuations contribute through an additive
linear random forcing of the free-surface evolution.
For deterministic resolved initial data, the solution is the
superposition of the deterministic Airy evolution and a randomly
forced linear wave response.
At this order, small-scale stochastic perturbations therefore
generate additional linear waves without nonlinear energy transfer
through advection of the resolved scales by the unresolved motion.
\paragraph{Time normalization for the nonlinear shallow-water models.}
For the remaining subsections we use the long-wave time $T=L/\sqrt{gH}$ and potential scale $\Phi_0=\epsilon L\sqrt{gH}$. This amounts to choosing $\nu=1$ in the scale definitions, rather than discarding an $\mathcal O(\mu)$ correction in the earlier dispersion-based time unit. We retain $\m=\tau/T=r/\sqrt{\nu_\sigma}$ with this time normalization and set $a=\epsilon_\sigma\m$, $\alpha=a/\epsilon$. The scaled fluctuating potential is defined accordingly. All fields below are dimensionless.
\end{revision}

\begin{revision}
\subsubsection{Stochastic Saint-Venant equations}
Let $\mu\ll1$, $\epsilon=\mathcal O(1)$ and $a=\mathcal O(\epsilon)$, with both wavelength ranges shallow and uniformly controlled derivatives. For example, comparable amplitudes with $r=\mathcal O(1)$ give this balance. At leading order in $\mu$, the resolved equations are
\begin{align}
&\partial_t\overline\varphi+\eta+\frac\epsilon2|\nabla\overline\varphi|^2
+a\nabla\varphi^\tau\cdot\nabla\overline\varphi=0,
\label{eq-adim-SNSW}\\
&\partial_t\eta=-\nabla\cdot(h\nabla\overline\varphi)-\alpha\nabla\cdot(h\nabla\varphi^\tau).
\label{SV-eta}
\end{align}
Taking the leading order of \eqref{varphi-i-final-adim} gives
\begin{equation}
\partial_t\varphi_i+\epsilon\nabla\overline\varphi\cdot\nabla\varphi_i=0.
\label{SV-CF}
\end{equation}
Thus the potential modes are transported by the resolved flow. This includes random sweeping and allows their spatial gradients and local wavevectors to be deformed by gradients of that flow. The omitted vertical noise--noise coupling is higher order only under the stated joint long-wave and derivative assumptions.

\subsection{Fully nonlinear Serre--Green--Naghdi approximation}
\label{sec:SGN}
Retaining first-order dispersive terms at $\epsilon,a=\mathcal O(1)$, let
\begin{equation}
\mathbb U[h]=I+\mu\mathbb T[h],\qquad
\nabla f=\mathbb U[h]{\cal V}_\eta f+\mathcal O(\mu^2),
\quad
{\cal C}_h(V)=-\nabla\cdot(hV)+\epsilon\nabla\eta\cdot\mathbb U[h]V.
\label{SGN-operators}
\end{equation}
With errors of formal order $\mathcal O(\mu^2)$ suppressed, the momentum and elevation equations are
\begin{align}
\partial_t(\mathbb U[h]\overline V)+\nabla\eta
+\frac\epsilon2\nabla|\mathbb U[h]\overline V|^2
+a\nabla(\mathbb U[h]V^\tau\cdot\mathbb U[h]\overline V)
+\mathbb Q(\overline V,V^\tau)&=0,
\label{eq-adim-SSGN}\\
\partial_t\eta&=-\nabla\cdot(h\overline V)-\alpha\nabla\cdot(hV^\tau),
\label{SGN-eta}
\end{align}
where
\begin{equation}
\mathbb Q(\overline V,V^\tau)
=-\frac{\mu\epsilon}{2}\nabla\bigl[{\cal C}_h(\overline V)^2\bigr]
-a\mu\nabla\bigl[{\cal C}_h(\overline V){\cal C}_h(V^\tau)\bigr].
\label{SGN-Q}
\end{equation}
The mode velocities are $V_i={\cal V}_\eta\varphi_i$. Taking the horizontal gradient of \eqref{varphi-i-final-adim} gives the consistent first-order mode equation
\begin{multline}
\partial_t(\mathbb U[h]V_i)
+\epsilon\nabla(\mathbb U[h]\overline V\cdot\mathbb U[h]V_i)\\
-\epsilon^2\mu\nabla\!\left[(\mathbb U[h]V_i\cdot\nabla\eta){\cal C}_h(\overline V)\right]
+a\mu\nabla\!\left[{\cal C}_h(V_i)\nabla\cdot(hV^\tau)\right]=0.
\label{SGN-mode}
\end{multline}
Every term is a horizontal vector; in particular, the third term includes the gradient of the full scalar product. The time derivative includes the time dependence of $h$. The depth-averaged velocities are recovered by inverting $\mathbb U[h]$ on an appropriate domain. Unlike the Saint-Venant reduction, the coupled mode family retains the stochastic vertical interaction. This formal truncation does not resolve the regularity issue of the full mode equations.

\subsection{Weakly nonlinear Boussinesq regime}
For $\epsilon=\mathcal O(\mu)$ and $a=\mathcal O(\epsilon)$, $\alpha=\mathcal O(1)$. To the retained order,
\[
\mathbb U[h]=\mathbb U_0+\mathcal O(\mu^2),\qquad
\mathbb U_0=I-\frac\mu3\nabla\nabla\cdot,
\]
and $\partial_t\mathbb U[h]=\mathcal O(\mu^2)$ for bounded elevation rates. Suppressing formal $\mathcal O(\mu^2)$ terms yields
\begin{align}
\mathbb U_0\partial_t\overline V+\nabla\eta
+\frac\epsilon2\nabla|\overline V|^2
+a\nabla(V^\tau\cdot\overline V)&=0,
\label{eq-adim-SBLN}\\
\partial_t\eta&=-\nabla\cdot(h\overline V)-\alpha\nabla\cdot(hV^\tau),
\label{Bouss-eta}\\
\mathbb U_0\partial_tV_i+\epsilon\nabla(\overline V\cdot V_i)&=0.
\label{Bouss-mode}
\end{align}
We use the same $\overline V,V^\tau,V_i$ throughout; no additional superscripted velocities are required. For irrotational velocities at this order, $\tfrac12\nabla|\overline V|^2=(\overline V\cdot\nabla)\overline V$. Thus the convective self-advection coefficient is $\epsilon$, not $\epsilon/2$. The gradient form also keeps the two symmetric cross-advection contributions together.

\subsection{Weakly nonlinear regime with stronger noise}
Consider the formal distinguished scaling $\epsilon=\mathcal O(\mu)$ and $a=\mathcal O(\epsilon/\sqrt\mu)=\mathcal O(\sqrt\mu)$. The stochastic dispersive terms of order $a\mu$ must then be retained. Using the same variables as above, a consistent truncation is
\begin{align}
\partial_t(\mathbb U[h]\overline V)+\nabla\eta
+\frac\epsilon2\nabla|\overline V|^2
+a\nabla(\mathbb U[h]V^\tau\cdot\mathbb U[h]\overline V)
+\mathbb Q_\sigma(\overline V,V^\tau)&=0,
\label{eq-adim-SBSN}\\
\partial_t\eta&=-\nabla\cdot(h\overline V)-\alpha\nabla\cdot(hV^\tau),
\label{strong-Bouss-eta}
\end{align}
where
\begin{equation}
\mathbb Q_\sigma(\overline V,V^\tau)
=-a\mu\nabla\!\left[(\nabla\cdot(h\overline V))(\nabla\cdot(hV^\tau))\right].
\end{equation}
The $\mathbb U[h]$ factors in the stochastic cross term retain corrections of the same order $a\mu$. The corresponding mode equation is
\begin{equation}
\partial_t(\mathbb U[h]V_i)+\epsilon\nabla(\overline V\cdot V_i)
-a\mu\nabla\!\left[(\nabla\cdot(hV_i))(\nabla\cdot(hV^\tau))\right]=0.
\label{strong-Bouss-mode}
\end{equation}
Here $\alpha=\mathcal O(\mu^{-1/2})$, so the stochastic elevation rate may be enhanced. In that case $\partial_t\mathbb U[h]$ can contribute at order $\mu^{3/2}$ and is retained inside the time derivatives. The weakly nonlinear elevation and derivative assumptions must remain valid over the interval considered; this stronger-noise regime does not provide a uniform long-time approximation. A spatially resolved implementation must also respect the wavelength restriction stated at the start of \S\ref{sec:reduced}.

Table~\ref{tab:model-hierarchy} summarizes the assumptions, scaling regimes and status of the stochastic formulation
and its reduced models.
\begin{table}
\centering
\caption{Hierarchy of the wave models and assumptions introduced at each
stage. Airy and shallow-water models are derived from the finite-correlation
closed system; the decorrelated model is a separate limiting branch.
Here $\mu=(H/L)^2$, $\epsilon=A/H$, $a=\epsilon_\sigma\mathcal m$
and $\mathcal m=\tau/T$; $a$ is the coefficient of stochastic advection
in the nondimensional resolved dynamic equation.}
\label{tab:model-hierarchy}
\begingroup
\small
\setlength{\tabcolsep}{3pt}
\renewcommand{\arraystretch}{1.16}
\begin{tabularx}{\textwidth}{@{}
>{\raggedright\arraybackslash}p{0.13\textwidth}
>{\raggedright\arraybackslash}X
>{\raggedright\arraybackslash}p{0.245\textwidth}
>{\raggedright\arraybackslash}p{0.075\textwidth}
>{\raggedright\arraybackslash}p{0.16\textwidth}@{}}
\toprule
Model & Representation and closure & Scaling regime & Time & Status \\
\midrule
Full system
& $\phi=\overline\phi+\phi^\tau$; all mixed and self-interactions
  retained. The quadratic fluctuation terms obstruct a direct
  white-noise formulation (zero-correlation). 
& Ideal potential flow; no small-amplitude or long-wave expansion.
& $\tau>0$
& Exact reformulation at finite $\tau$ within the potential-flow
  assumptions. \\
\addlinespace
Closed stochastic system
& LU surface balance $R_s\approx0$; Euler expectation principle
  for modes, with specified modal pressure boundary conditions.
& No small-amplitude or long-wave expansion.
& $\tau>0$
& Modeling closure; canonical resolved dynamics at fixed $s$. \\
\addlinespace
Decorrelated limit
& Stratonovich increments replace regularized temporal factors;
  modes remain dynamical.
& Fixed integrated noise covariance normalization;
  no long-wave expansion.
& $\tau\to0$
& Formal limit for the coupled wave--mode system. \\
\addlinespace
Airy
& Linearization about rest; additive surface forcing;
  $\partial_t\varphi_i=0$ at this order.
& $\epsilon\ll1$, $a=\mathcal O(\epsilon)$;
  full finite-depth dispersion may be retained.
& $\tau>0$
& Formal linear approximation of the closed system. \\
\addlinespace
Saint-Venant
& Leading long-wave terms; vertical mode--noise coupling omitted.
& $\mu\ll1$, $\epsilon=\mathcal O(1)$,
  $a=\mathcal O(\epsilon)$.
& $\tau>0$
& Formal leading-order shallow-water approximation. \\
\addlinespace
SGN
& First-order dispersive corrections retained,
  including mode--noise interactions.
& $\mu\ll1$, $\epsilon,a=\mathcal O(1)$;
  terms through first order in $\mu$ retained.
& $\tau>0$
& Formal dispersive shallow-water approximation. \\
\addlinespace
Boussinesq
& Weakly nonlinear dispersive reduction;
  stronger-noise variant specified.
& $\mu\ll1$, $\epsilon=\mathcal O(\mu)$,
  $a=\mathcal O(\epsilon)$ in the standard regime.
& $\tau>0$
& Formal weakly nonlinear, dispersive approximation. \\
\bottomrule
\end{tabularx}
\endgroup

\medskip
\begin{minipage}{\textwidth}
\footnotesize
\textit{Closure convention.}
$s=\varphi^\tau$ and $R_s=\partial_t s+B_\eta(s)$.
The notation $R_s\approx0$ specifies the fluctuation block omitted
when defining the resolved model; the separately reconstructed noise
is governed by its mode dynamics. The canonical statement concerns
the resolved variables for a prescribed fluctuating trace.

\smallskip
\textit{Common long-wave assumptions.}
The Saint-Venant, Serre--Green--Naghdi and Boussinesq rows require
both spatial scale ranges to be shallow:
$\mu\ll1$ and $\mu_\sigma=\mu/r^2\ll1$, where $r=L_\sigma/L$,
together with positive depth and appropriately controlled spatial
derivatives. The Airy linearization itself requires no expansion in $\mu$.

\smallskip
\textit{Stronger-noise Boussinesq variant.}
With $\epsilon=\mathcal O(\mu)$ and $a=\mathcal O(\sqrt\mu)$,
the additional stochastic dispersive terms of order
$a\mu=\mathcal O(\mu^{3/2})$ are retained.
The reduced systems listed here are derived at finite correlation
time; their white-noise versions involve a further limiting step.
\end{minipage}
\end{table}
\end{revision}

\begin{revision}
\section{Surface transport, refraction and temporal regularization}
\label{sec:GO}
\subsection{Geometric optics for the correlation modes}
We use dimensional variables and retain terms of first order
in the noise amplitude, with the spatial scales held fixed.
The mode--noise interaction $B_iG[\eta]\varphi^\tau$ in
\eqref{varphi-i-sto} is quadratic in the noise amplitude
and is therefore discarded at this order.
The correlation modes are then transported by the resolved
surface velocity:
\begin{equation}
\partial_t\varphi_i+u\cdot\nabla\varphi_i=0,
\qquad u=\left.\nabla_x\overline\phi\right|_\eta
=\nabla\overline\varphi-\overline B\nabla\eta.
\label{GO-trace-transport}
\end{equation}
This calculation requires no expansion in $\mu$.
For a deep-water realization at a flat reference surface $z=0$,
the decaying harmonic extension satisfies
\[
\widehat{\phi_i}(\xi,z)
=e^{|\xi|z}\widehat{\varphi_i}(\xi),
\qquad
\left.\partial_z\widehat{\phi_i}\right|_0
=|\xi|\widehat{\varphi_i}.
\]
The same leading surface-transport law applies at finite and
infinite depth. The depth enters through the harmonic extensions
and the resolved flow.

Let us use the WKB ansatz
\begin{equation}
\varphi_j(x,t)=b_j(x,t)\exp\!\left(\frac{i{\cal P}_j(x,t)}{\delta}\right)+\mathrm{c.c.},
\qquad 0<\delta\ll1,
\label{WKB-ansatz}
\end{equation}
where $b_j$ and the resolved velocity vary slowly relative to the oscillations, and $\delta$ is distinct from the wave-amplitude parameter $\epsilon$. Define the scaled phase gradient and frequency by
\begin{equation}
k_j=\nabla{\cal P}_j,\qquad \omega_j=-\partial_t{\cal P}_j;
\label{phase-der}
\end{equation}
the physical oscillatory wavenumber and frequency carry the common factor $\delta^{-1}$. Inserting \eqref{WKB-ansatz} into \eqref{GO-trace-transport}, the order $\delta^{-1}$ equation is the real Hamilton--Jacobi equation
\begin{equation}
\partial_t{\cal P}_j+u\cdot\nabla{\cal P}_j=0,
\qquad \omega(x,k,t)=k\cdot u(x,t).
\label{GO-Hamilton-Jacobi}
\end{equation}
The characteristic equations, in position--wavevector order, are
\begin{equation}
\frac{\dif x_j}{\dif t}=\nabla_k\omega=u(x_j,t),
\qquad
\frac{\dif k_j}{\dif t}=-\nabla_x\omega=-\bigl(\nabla u(x_j,t)\bigr)^{\mathsf T}k_j.
\label{WL}
\end{equation}
At the next order the scalar envelope is transported:
\begin{equation}
\partial_tb_j+u\cdot\nabla b_j=0.
\label{GO-envelope}
\end{equation}
Thus the correlation patterns are swept by the resolved surface
velocity, while their wavevectors evolve through rotation and
straining by its spatial gradient.
This is consistent with random sweeping
\cite{Chen-Kraichnan-1989,Kraichnan-1964,Tennekes-1975}
and with the kinematics of rapid distortion.
The scalar envelope remains constant along the surface
characteristics.
The Hamiltonian $\omega=k\cdot u$ describes the advection and
refraction of the correlation patterns, while intrinsic
gravity-wave dispersion belongs to the resolved free-surface
dynamics.
\paragraph{Connection with wave--current interactions.}
The stochastic transport terms admit a direct wave--current
interpretation. At finite correlation time, the unresolved
advecting velocity $u^\tau$ acts as a random current.
When this velocity is approximately uniform over a wave packet
and its penetration depth, the gravity-wave frequency takes
the familiar Doppler-shifted form
\begin{equation}
\omega(x,k,t)=\omega_0(k)+k\cdot u^\tau(x,t),
\qquad
\omega_0(k)=\sqrt{g|k|\tanh(H|k|)}.
\label{wave-current-doppler}
\end{equation}
In deep water, $\omega_0(k)=\sqrt{g|k|}$.
Spatial variations of the advecting velocity produce random
refraction and redistribution across wavevectors, connecting
the stochastic transport mechanism with current-induced
wave variability
\cite{Ardhuin2017SmallScale,VillasBoas2020WaveCurrent}.
In the complementary direction, \eqref{WL} describes transport
and refraction of the correlation patterns by the resolved
surface velocity.
The coupled formulation thus provides a dynamical representation
of unresolved wave--current transport within the present
potential-flow framework.

\end{revision}

\begin{revision}
\subsection{A causal Ornstein--Uhlenbeck representation}
\label{sec:OU}
An alternative temporal regularization uses the causal, unit-integral kernel
\begin{equation}
h_\tau(t)=\frac1\tau e^{-t/\tau}\mathbf1_{t\geq0},\qquad
Z_t^{\tau,i}=\frac1\tau\int_{-\infty}^{t}e^{-(t-s)/\tau}\dif\beta_s^i,
\qquad \varphi^\tau=\sum_i\varphi_iZ_t^{\tau,i}.
\label{OU-Ito-regul-noise}
\end{equation}
This kernel is causal and is neither symmetric nor compactly supported, so it extends the initial kernel class. The stationary temporal factors satisfy
\begin{align}
&Z_t^{\tau,i}=e^{-1}Z_{t-\tau}^{\tau,i}
+\frac1\tau\int_{t-\tau}^{t}e^{-(t-s)/\tau}\dif\beta_s^i,
\label{OU-recursion}\\
&\dif Z_t^{\tau,i}=-\frac1\tau Z_t^{\tau,i}\dif t+\frac1\tau\dif\beta_t^i,
\label{OU-SDE}\\
&\mathbb E[Z_t^{\tau,i}Z_s^{\tau,j}]=\frac{\delta_{ij}}{2\tau}e^{-|t-s|/\tau}.
\label{OU-covariance}
\end{align}
In particular, the variance is $1/(2\tau)$, consistently with the order $\Upsilon_\tau\sim\tau^{-1}$ in \eqref{var}. The diffusion coefficient in \eqref{OU-SDE} is $1/\tau$.

For finite $\tau$, sufficiently regular solutions of the mode equation have finite variation in time. The product rule therefore gives
\begin{equation}
\dif\varphi^\tau=\sum_iZ_t^{\tau,i}\dif\varphi_i
-\frac1\tau\varphi^\tau\dif t+\frac1\tau\sum_i\varphi_i\dif\beta_t^i,
\label{OU-potential-product}
\end{equation}
with zero quadratic covariation $\langle\varphi_i,Z^{\tau,i}\rangle_t$. If the spatial modes are frozen, $\dif\varphi_i=0$, this reduces to a linear OU field. With evolving random modes, only the temporal factors are OU process and the full fluctuating potential is even possibly  nonGaussian.

The normalization can also be checked by integration:
\begin{equation}
\int_0^t Z_s^{\tau,i}\dif s=\beta_t^i-\beta_0^i-\tau(Z_t^{\tau,i}-Z_0^{\tau,i}).
\label{OU-integrated}
\end{equation}
The last term vanishes in probability uniformly on compact time intervals as $\tau\to0$. Integrated independent OU factors therefore approximate Brownian motion. Under the usual smooth-coefficient assumptions, their finite-dimensional Wong--Zakai limit is Stratonovich -- see \cite[\S11.7.3]{PavliotisStuart2008}. This fact motivates, but does not prove, the formal coupled SPDE limit in \S\ref{sec:final-models}. With a diffusion coefficient of order $\tau^{-1/2}$ instead, the integrated noise would vanish in this limit. Finite-memory OU forcing remains useful in ocean and climate modeling \cite{Hasselmann-1976,McWilliams-Huckle-2006,Saravanan-McWiliams-1998} independently of a white-noise approximation.
\end{revision}

\section{Numerical assessment: wave deformation and a filtered SGN reference}
\label{sec:numerical-sv}
We present two preliminary assessments of the one-dimensional stochastic
Saint--Venant (SV) model: phase-resolved illustrations of mode refraction
and wave deformation, followed by a comparison with a filtered deterministic
Serre--Green--Naghdi (SGN) ensemble. The mechanism figures are unfiltered;
the distributional comparison retains only $|k|\leq3$.

\paragraph{Model and numerical setup.}
On the periodic domain $x\in[0,2\pi)$, let
$u=\partial_x\overline\varphi$, $h=1+\epsilon\eta$, and
$v=\sum_i\partial_x\varphi_iW_t^{\tau,i}$. We integrate
\begin{equation}
\begin{aligned}
&\partial_t\eta+\partial_x[h(u+\alpha v)]=0,\\
&\partial_tu+\partial_x[\eta+\epsilon u^2/2+\epsilon\alpha uv]=0,\\
&\partial_t\varphi_i+\epsilon u\,\partial_x\varphi_i=0,
\qquad \alpha=\epsilon_\sigma\mathcal m/\epsilon,
\quad \mathcal m=\tau/T.
\end{aligned}
\label{eq:numerical-sv}
\end{equation}
The modes evolve directly, without a WKB approximation.
Temporal factors use the symmetric unit-integral kernel
$h_\tau(t)=\tau^{-1}b(t/\tau)$, with
$b(q)\propto\exp[-1/(1-q^2)]$ on $|q|<1$ and zero elsewhere.
Their variance is $I_b/\tau$, where $I_b=\int b^2$.
The mechanism experiments use $\epsilon=0.15$, $\mathcal m=0.2$,
and six initial modes
$\cos(kx)/(k\sqrt3)$ and $\sin(kx)/(k\sqrt3)$, $k=4,5,6$,
whose squared gradients sum to one.

\paragraph{Mode transport, steepening and refraction.}
We compare deterministic SV with frozen modes, modes transported by a
prescribed ensemble-mean velocity, and fully coupled modes.
The mean velocity is estimated from an ensemble mean.
Stochastic evaluation cases share the same noise paths.

A 192-member sinusoidal experiment,
$\eta(x,0)=u(x,0)=0.6\cos x$, uses for different values for 
$\alpha$: $0.0075$, $0.015$ , $0.0225$ and $0.03$.
Over this range, the mean paired elevation RMS differences between
coupled and frozen transport, and between coupled and mean transport,
grow approximately as $\alpha$ and $\alpha^2$, respectively.
This identifies the additional realization-dependent feedback supplied
by fully coupled mode evolution.

\begin{figure}[!ht]
 \centering
 \includegraphics[width=\linewidth]{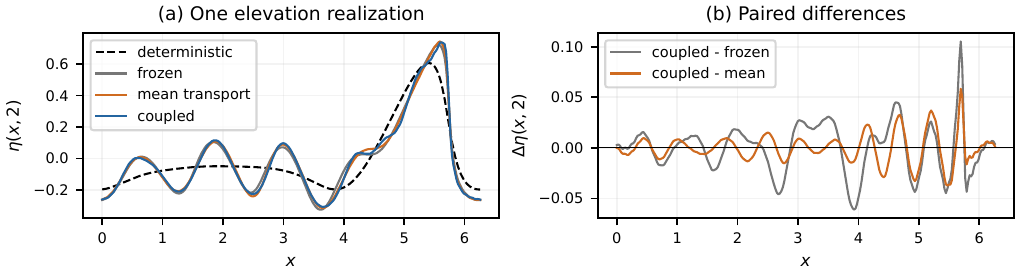}
 \caption{First evaluation realization of the SV packet at $t=2$,
 $\alpha=0.03$. (a) Deterministic and matched-noise stochastic elevations.
 (b) Magnified paired differences. The unfiltered profiles show local
 steepening, additional short waves and the effect of evolving modes.}
 \label{fig:numerical-realisation}
\end{figure}

For the localized packet, we take
$\eta=u=0.6(q-\langle q\rangle)/\|q-\langle q\rangle\|_\infty$,
where $q=e^{2(\cos(x-\pi)-1)}\cos(2(x-\pi))$.
Figure~\ref{fig:numerical-realisation} shows the first evaluation
realization. At $t=2$ and $\alpha=0.03$, the maximum elevation gradient
increases from 2.36 for deterministic SV to 8.77 for coupled SV,
and the crest increases from 0.606 to 0.739.
These sample values illustrate local steepening and crest modulation.

\begin{figure}[!ht]
 \centering
 \includegraphics[width=\linewidth]{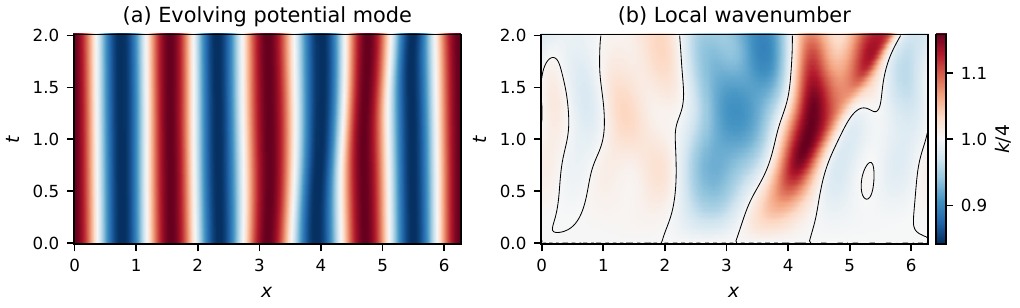}
 \caption{Mode refraction in the first fine-grid packet realization
 at $\alpha=0.015$. (a) Transported initially $k=4$ potential mode.
 (b) Normalized local wavenumber; the black contour marks $k/4=1$.
 The phase-gradient diagnostic requires no WKB approximation.}
 \label{fig:numerical-refraction}
\end{figure}

For an advected sine--cosine pair $(c,s)$, the local wavenumber is
$k=(c\,\partial_xs-s\,\partial_xc)/(c^2+s^2)$ and satisfies
$\partial_tk+\partial_x(\epsilon uk)=0$.
At $\alpha=0.015$, the normalized wavenumber $k/4$ ranges from
0.894 to 1.158, directly identifying extension and compression
of the modes (Fig.~\ref{fig:numerical-refraction}).

These runs use Fourier projection $|k|<N_x/4$ and fourth-order
Runge--Kutta with noise evaluated at intermediate stages.
Shared-path, 96-member refinements from $(N_x,\Delta t)=(128,0.004)$
to $(256,0.002)$ change the mean paired RMS metrics by at most 0.14\%,
with positive depth throughout the checks.

\paragraph{Comparison with a filtered SGN ensemble.}
The reference solves the classical flat-bottom SGN equations,
with depth-averaged velocity $U$:
\begin{equation}
 \eta_t=-(hU)_x,\qquad
 U_t+\epsilon UU_x+\eta_x
 =\frac{\mu}{3h}\partial_x\!\left[h^3
 \bigl(U_{xt}+\epsilon(UU_{xx}-U_x^2)\bigr)\right].
 \label{eq:numerical-sgn-reference}
\end{equation}
We use $\epsilon=0.15$, $\mu=0.0025$ and $t\leq2$.
The initial long packet is projected to $|k|\leq3$.
Reference members add Gaussian velocity coefficients at $k=4,5,6$,
with variance $\sigma^2/3$ per quadrature for $\sigma=0.06,0.12$.
Their elevations satisfy the right-going linear SGN relation
$U_k=c_k\eta_k$, $c_k=(1+\mu k^2/3)^{-1/2}$.
Both coarse models thus start from identical deterministic resolved
states, with reference uncertainty confined initially to the short band.

An independent 64-member SGN pilot calibrates the initial short-band
velocity covariance and its positive-lobe correlation integral $T_I$,
taken up to the first zero of the normalized Eulerian autocorrelation.
The mode shapes match this covariance, with
$\tau=2I_bT_I$ and $\alpha=\widehat\sigma\sqrt{\tau/I_b}$.
This gives $\tau\simeq0.274$ and $\alpha=0.0411,0.0821$,
without tuning on the test ensemble.

The held-out comparison uses 128 reference and 128 stochastic SV members.
SGN uses $N_x=256$, $\Delta t=0.002$, Fourier projection
$|k|<N_x/6$ and an elliptic solve at each Runge--Kutta stage.
SV uses $N_x=128$ for its auxiliary modes and projects
both surface right-hand sides to $|k|\leq3$ at every stage.
Reference outputs receive the same filter.
We score elevation using sliced Wasserstein-1 distance $\mathcal D_3$,
with 128 fixed unit-direction projections of
$\sqrt2(\Re\widehat\eta_{1:3},\Im\widehat\eta_{1:3})$,
where $\widehat\eta_k=(2\pi)^{-1}\int_0^{2\pi}\eta e^{-ikx}\,\mathrm dx$.

\begin{figure}[!ht]
 \centering
 \includegraphics[width=\linewidth]{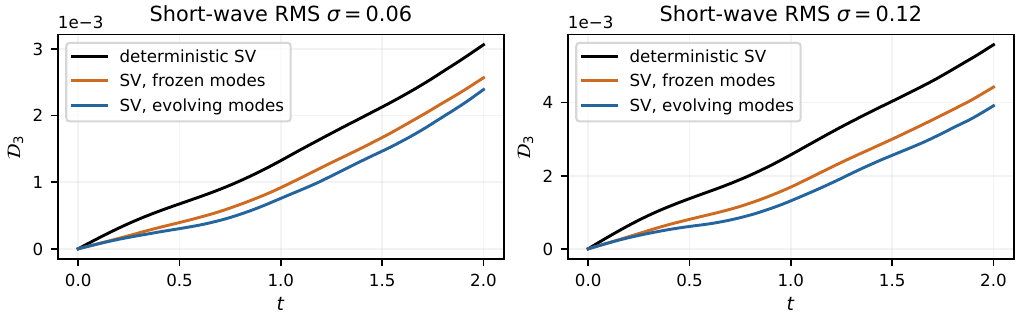}
 \caption{Distance to the filtered held-out SGN elevation distribution
 for the two uncertainty amplitudes. All scored fields retain
 $|k|\leq3$. Coupled evolving modes improve the score over deterministic
 SV and frozen modes over most of the interval.}
 \label{fig:numerical-reference}
\end{figure}

At $t=2$, coupled SV reduces the distance relative to deterministic
SV by 21.9\% and 29.9\% for $\sigma=0.06$ and $0.12$, respectively;
the time-integrated reductions are 35.2\% and 41.0\%.
Evolving modes also improve on frozen modes
(Fig.~\ref{fig:numerical-reference}).
The improvement is partial: stochastic SV reproduces only
17.3\% and 16.5\% of the reference variance, and its mean errors, 0.00573 and
0.00747, slightly exceed deterministic errors 0.00552 and 0.00710. The Wasserstein distance of the stochastic SV with  the reference distribution is nevertheless smaller.

These preliminary results support a partial correction
of the low-bandwidth elevation distribution, together with phase-resolved
steepening and refraction. Quantitative variance closure and assessment
against fully nonlinear water waves remain for further work.

\begin{revision}
\section{Conclusion}
In this paper, we have developed a stochastic variational framework for surface gravity waves that connects phase-resolved dynamics with a statistical representation of unresolved scales. Starting from Luke's classical variational principle for
irrotational incompressible free-surface flow, we introduced a decomposition of the velocity potential into a resolved
large-scale component and a regularized stochastic component. Combined with a pathwise variational principle and a closure
of the fast surface-fluctuation balance, this decomposition yields stochastic free-surface evolution coupled to harmonic
bulk potentials. For a prescribed fluctuating trace, the resolved system admits a non-autonomous canonical Hamiltonian formulation of Zakharov--Craig--Sulem type.

To close the system and describe the dynamics of the unresolved scales, we use a second variational principle in expectation
for the noise correlation modes. Following its Euler formulation in \cite{Debussche-Memin25}, we express the resulting linearized mode dynamics in potential-flow form. The resulting coupled system, combining large-scale dynamics with correlation-mode evolution, provides a systematic framework for deriving stochastic models of ocean waves.
At leading order in the noise amplitude, a WKB analysis yields a Hamilton--Jacobi equation and associated ray-tracing dynamics. The correlation patterns are randomly swept by the resolved surface velocity, while their wavevectors are rotated and strained by its spatial gradient, connecting the formulation with geometric optics, random sweeping and rapid-distortion kinematics.

Small-amplitude and long-wave asymptotic expansions then yield simplified models for several physical regimes. In the long-wave, small-amplitude limit, the theory reduces to a stochastic Airy-wave model with additive random forcing of the free-surface evolution. For deterministic resolved initial data, its solution is the superposition of a deterministic carrier and a randomly forced linear wave response. In shallow water with stronger nonlinearity, stochastic Saint-Venant equations emerge with transport noise representing unresolved advection. Retaining dispersive corrections yields stochastic Serre--Green--Naghdi and Boussinesq-type systems, with the noise scaling determining which interaction terms must be retained. A causal Ornstein--Uhlenbeck representation provides a simple colored-noise temporal regularization that can be combined with the evolving spatial correlation modes.

Several consequences follow from this formulation. The variational structure constrains the resolved--noise
interactions and provides an explicit energy budget. For smooth finite-correlation solutions, the total-energy budget
quantifies the energy exchange associated with departures from the adopted surface fluctuation balance.
When this balance is satisfied exactly, the total wave energy is conserved. The framework also offers a route toward simplified stochastic surface-wave models and their coupling with mean currents. The geometric-optics description suggests numerical strategies combining evolving correlation modes with ray-tracing and spectral methods. \red{Preliminary simulations of the stochastic Saint-Venant system against a filtered Serre--Green--Naghdi reference show that the modes are transported and deformed by the resolved flow, and that the noise partially corrects the distribution of the resolved elevation, although the reference variance is only partly reproduced.}

The companion paper \cite{DebusscheMemin2026b} develops the
kinetic implications and geophysical consequences further,
including comparisons of transport-induced and classical
four-wave interaction rates using JONSWAP spectra.
These comparisons suggest regimes in which stochastic transport
by unresolved scales may contribute significantly alongside
classical wave-interaction mechanisms, motivating further
assessment of transport contributions to spectral-wave closures.

Several directions remain for future work.
Mathematical questions include the well-posedness of the coupled
system under appropriate spatial regularity conditions on the
noise modes, the vanishing-correlation-time limit of this system,
and quantitative justification of the asymptotic and
geometric-optics approximations.
Physical extensions include broader turbulence spectra,
wave breaking, surf-zone dynamics, wind input and dissipation,
together with their associated energy budgets.
\red{Systematic numerical assessment of the full models in
phase-resolved and spectral settings remains an essential step
toward applications.}

By coupling the resolved wave field to dynamically evolving
correlation modes, this framework makes unresolved spatial
structure part of the model dynamics.
Combining canonical resolved dynamics, multiscale asymptotics
and stochastic mode evolution provides a foundation for
water-wave models that represent transport, uncertainty
and interactions across scales.

\end{revision}

\paragraph*{Acknowledgements:} 
The authors acknowledge the support of the ERC EU project 856408-STUOD.

\begin{appendix}

\begin{revision}
\section{Dirichlet-to-Neumann shape derivative}
\label{DTN-O}
We recall the standard identities needed for the variational derivatives; see \cite[Chapter~3]{Lannes-2013}. The operator $G[\eta]$ is defined by harmonic extension as in \eqref{def-DtoN}, with a flat-bottom homogeneous Neumann condition and periodic horizontal boundaries or sufficient decay. For a fixed surface trace $f$ and an elevation perturbation $\zeta$, let
\begin{equation}
B_f=\frac{G[\eta]f+\nabla\eta\cdot\nabla f}{1+|\nabla\eta|^2},
\qquad U_f=\nabla f-B_f\nabla\eta.
\end{equation}
The shape derivative, taken at fixed $f$, is
\begin{equation}
\mathrm D_\eta G[\eta]\cdot\zeta\, f
=-G[\eta](\zeta B_f)-\nabla\cdot(\zeta U_f).
\label{DTN-shape}
\end{equation}
Self-adjointness and integration by parts then give the variational derivative of the integrated quadratic form:
\begin{equation}
\frac\delta{\delta\eta}\left(\frac12\langle f,G[\eta]f\rangle\right)
=\frac12|\nabla f|^2
-\frac12\frac{(G[\eta]f+\nabla\eta\cdot\nabla f)^2}{1+|\nabla\eta|^2}
=B_\eta(f).
\label{DTN-quadratic}
\end{equation}
The derivative is a density with respect to the horizontal pairing. Polarization yields the cross-energy derivative used in Appendix~B.
\end{revision}

\section{Hamiltonian reduction and energy balance}

\begin{revision}
The fluctuating trace is held fixed in variations of the resolved variables. Let $p=\overline\varphi$, $s=\varphi^\tau$ and $\langle f,k\rangle=\int_{{\cal D}^h}fk\,\dif x$; all variables in this appendix are dimensional.

\subsection{Hamiltonian definition and relation to Luke's action}
\label{Reduction-Luke}
The definition of the resolved model is the canonical action
\begin{equation}
\mathcal S_H=\int_{t_0}^{t_1}[\langle p,\partial_t\eta\rangle-\overline{\mathcal H}(\eta,p;s)]\dif t,
\label{app-canonical-action}
\end{equation}
where
\begin{equation}
\overline{\mathcal H}(\eta,p;s)=\tfrac12\langle p,G[\eta]p\rangle
+\langle s,G[\eta]p\rangle+\tfrac g2\|\eta\|_{L^2}^2.
\label{app-H}
\end{equation}
This definition imposes no Bernoulli equation on $s$. To relate it to Luke's action, harmonic extension and time integration by parts give, up to endpoint terms and a constant,
\begin{equation}
\frac{\mathcal S_{\rm Luke}}\rho
=\mathcal S_H+\int_{t_0}^{t_1}\left[\langle s,\partial_t\eta\rangle-\tfrac12\langle s,G[\eta]s\rangle\right]\dif t.
\label{app-reduced-action-bulk}
\end{equation}
Here Green's identity gives
\begin{equation}
\int_{\Omega_\eta}\nabla\overline\phi\cdot\nabla\phi^\tau\dif x\,\dif z
=\langle s,G[\eta]p\rangle=\langle p,G[\eta]s\rangle.
\label{app-cross-energy}
\end{equation}
The extra terms in \eqref{app-reduced-action-bulk} are independent of $p$. At fixed $s$, their variation in the direction $\zeta=\delta\eta$, vanishing at the temporal endpoints, is
\begin{equation}
\delta_\eta\left(\frac{\mathcal S_{\rm Luke}}\rho-\mathcal S_H\right)\cdot\zeta
=-\int_{t_0}^{t_1}\langle\partial_ts+B_\eta(s),\zeta\rangle\dif t.
\label{Luke-action-residual}
\end{equation}
Thus the resolved Hamiltonian equations coincide with the reduced Luke equations when $\partial_ts+B_\eta(s)=0$. Otherwise they define a closure with the stated residual. 
\end{revision}

\subsection{Canonical structure}

For a prescribed fluctuating trace $s(x,t)$, the canonical term
$\langle p,\partial_t\eta\rangle$ in the resolved action identifies
$\eta$ and $p$ as conjugate variables.
For sufficiently smooth, possibly time-dependent functionals
$F[\eta,p;t]$ and $K[\eta,p;t]$, the corresponding Poisson
bracket is
\begin{equation}
\{F,K\}
=
\int_{{\cal D}^h}
\left(
\frac{\delta F}{\delta\eta}\frac{\delta K}{\delta p}
-\frac{\delta K}{\delta\eta}\frac{\delta F}{\delta p}
\right)\dif x,
\label{app-bracket}
\end{equation}
where all variations are taken at fixed $s$.
This bracket generates the evolution of an observable through
\[
\frac{\dif F}{\dif t}
=\{F,\overline{\mathcal H}\}
+\left.\frac{\partial F}{\partial t}\right|_{\eta,p}.
\]
The last term accounts for explicit time dependence, including
any dependence on the prescribed trace $s(x,t)$.
In particular, it gives
$\partial_t\eta=\delta\overline{\mathcal H}/\delta p$
and
$\partial_t p=-\delta\overline{\mathcal H}/\delta\eta$.

The bracket has the same canonical form as that of the
total-potential water-wave system in $(\eta,\varphi)$.
Indeed, at each fixed time, the translation
$p=\varphi-s$ satisfies $\delta p=\delta\varphi$ when $s$
is prescribed.
The closure enters the Hamiltonian
$\overline{\mathcal H}$, while the canonical pairing is preserved.
This structure concerns the resolved variables while a Poisson
structure for the augmented system would additionally involve
the correlation modes and their couplings.

For a surface potential \(f\), denote its kinetic Bernoulli term by
\begin{equation}
B_\eta(f)
=
\frac12|\nabla f|^2
-\frac12\frac{(G[\eta]f+\nabla\eta\bcdot\nabla f)^2}
{1+|\nabla\eta|^2}.
\label{app-noise-energy-derivative}
\end{equation}
The shape-derivative identity in Appendix A gives
\[
\frac{\delta\overline{\mathcal H}}{\delta p}
=G[\eta](p+s),
\qquad
\frac{\delta\overline{\mathcal H}}{\delta\eta}
=g\eta+B_\eta(p+s)-B_\eta(s).
\]
Hamilton's equations are therefore
\begin{align}
\partial_t\eta
&=G[\eta](p+s),
\label{app-Ham-eta}\\
\partial_t p
&=-g\eta-B_\eta(p+s)+B_\eta(s).
\label{app-Ham-varphi}
\end{align}
The second equation expands to \eqref{eq-S-varphi}, while the first
is the kinematic boundary condition \eqref{eq-S-dyn-bound-cond}.

\subsection{Energy balance at finite correlation time}
\label{app-energy-balance}

For smooth finite-correlation solutions with no lateral energy flux,
the total energy per unit density is
\begin{equation}
\mathcal H_{\rm tot}
=\overline{\mathcal H}(\eta,p;s)+\mathcal H_s(\eta,s)
=\frac12\langle p+s,G[\eta](p+s)\rangle
+\frac{g}{2}\|\eta\|_{L^2}^2,
\qquad
\mathcal H_s=\frac12\langle s,G[\eta]s\rangle.
\label{app-total-energy}
\end{equation}
Its fluctuation self-energy satisfies
\(\delta_s\mathcal H_s=G[\eta]s\) and
\(\delta_\eta\mathcal H_s=B_\eta(s)\).
In differentiating the interaction energy, the time dependence of the
Dirichlet-to-Neumann operator is included explicitly:
\[
\frac{\dif}{\dif t}\langle s,G[\eta]p\rangle
=
\langle\partial_t s,G[\eta]p\rangle
+\langle s,G[\eta]\partial_t p\rangle
+\langle s,(\mathrm D_\eta G[\eta]\bcdot\partial_t\eta)p\rangle.
\]
The last two terms enter the variational derivatives of
\(\overline{\mathcal H}\) with respect to \(p\) and \(\eta\).
By \eqref{app-Ham-eta}--\eqref{app-Ham-varphi}, their canonical
contributions cancel, leaving
\[
\frac{\dif\overline{\mathcal H}}{\dif t}
=
\left\langle\frac{\delta\overline{\mathcal H}}{\delta p},
\partial_t p\right\rangle
+
\left\langle\frac{\delta\overline{\mathcal H}}{\delta\eta},
\partial_t\eta\right\rangle
+\langle G[\eta]p,\partial_t s\rangle
=\langle G[\eta]p,\partial_t s\rangle.
\]
On the other hand,
\[
\frac{\dif\mathcal H_s}{\dif t}
=
\langle G[\eta]s,\partial_t s\rangle
+\langle B_\eta(s),G[\eta](p+s)\rangle.
\]
Consequently,
\begin{equation}
\frac{\dif\mathcal H_{\rm tot}}{\dif t}
=
\bigl\langle G[\eta](p+s),\,
\partial_t s+B_\eta(s)\bigr\rangle.
\label{app-mode-energy-residual}
\end{equation}

\begin{revision}
This is an energy budget: the right-hand side is the power associated with the fluctuating component and the closure residual. It vanishes if $\partial_ts+B_\eta(s)=0$, in which case the total energy is conserved pathwise and, under integrability, in expectation. For $s$ reconstructed from \eqref{W-v} and the mode dynamics, that balance is not generally satisfied and need not be a small error. The identity measures the resulting exchange without asserting conservation. Conversely, vanishing of the integrated power alone does not imply the pointwise fluctuation balance.

For the OU temporal factors of \S\ref{sec:OU}, $s$ is a semimartingale rather than differentiable. Since the finite-$\tau$ resolved variables have finite variation, the corresponding finite-$\tau$ identity takes the stochastic form
\begin{equation}
\dif\mathcal H_{\rm tot}
=\langle G[\eta](p+s),\circ\dif s\rangle
+\langle B_\eta(s),G[\eta](p+s)\rangle\dif t.
\label{OU-energy-budget}
\end{equation}
Its It\^o expression includes the quadratic variation of $s$ in the fluctuation self-energy. This finite-correlation identity should not be identified with an energy law for the formal white-noise model, whose fluctuation self-energy requires a separate treatment.
\end{revision}

\end{appendix}


\begingroup
\small
\begin{thebibliography}{10}

\bibitem{Agresti-et-al-2022}
Antonio Agresti, Matthias Hieber, Amru Hussein, and Martin Saal.
\newblock The stochastic primitive equations with transport noise and turbulent
  pressure.
\newblock {\em Stochastics and Partial Differential Equations: Analysis and
  Computations}, 12(1):53--133, 2024.
  
  \bibitem{AndrewsMcIntyre1978GLM}
David~G. Andrews and Michael~E. McIntyre.
\newblock {An exact theory of nonlinear waves on a Lagrangian-mean flow}.
\newblock {\em Journal of Fluid Mechanics}, 89(4):609--646, December 1978.

\bibitem{Airy1845}
George~Biddell Airy.
\newblock Tides and waves.
\newblock {\em Encyclopaedia Metropolitana}, 5:241--396, 1845.

\bibitem{Apolinario-et-al-23}
Gabriel~B. Apolinario, Geoffrey Beck, Laurent Chevillard, Isabelle Gallagher,
  and Ricardo Grande.
\newblock A linear stochastic model of turbulent cascades and fractional
  fields.
\newblock {\em Ann. Sc. Norm. Super. Pisa Cl. Sci}, 26(4):2043--2103, 2025.

\bibitem{Ardhuin2017SmallScale}
Fabrice Ardhuin, Sarah~T. Gille, Dimitris Menemenlis, Cesar~B. Rocha,
Nicolas Rascle, Bertrand Chapron, Jonathan Gula, and Jeroen Molemaker.
\newblock Small-scale open ocean currents have large effects
on wind wave heights.
\newblock \emph{Journal of Geophysical Research: Oceans},
122(6):4500--4517, 2017.
\newblock doi:10.1002/2016JC012413.

\bibitem{Bauer-et-al-JPO-20}
Werner Bauer, Pranav Chandramouli, Bertrand Chapron, Long Li, and Etienne
  M{\'e}min.
\newblock Deciphering the role of small-scale inhomogeneity on geophysical flow
  structuration: A stochastic approach.
\newblock {\em Journal of Physical Oceanography}, 50(4):983 -- 1003, 01 Apr.
  2020.
\bibitem{BoulvardMeminSainteMarie2026}
Pierre-Marie Boulvard, Etienne M{\'e}min, and Jacques Sainte-Marie.
\newblock Discrete numerical schemes for stochastic shallow water models
under location uncertainty.
\newblock \emph{Journal of Computational Physics}, 559:114913, 2026.
\newblock \href{https://doi.org/10.1016/j.jcp.2026.114913}
{doi:10.1016/j.jcp.2026.114913}.

\bibitem{Brzezniak-Slavik-2021}
Zdzis{\l}aw Brze{\'z}niak and Jakub Slav{\'\i}k.
\newblock Well-posedness of the 3{D} stochastic primitive equations with
  multiplicative and transport noise.
\newblock {\em Journal of Differential Equations}, 296:617--676, 2021.

\bibitem{Carigi-Luongo-2023}
Giulia Carigi and Eliseo Luongo.
\newblock Dissipation properties of transport noise in the two-layer
  quasi-geostrophic model.
\newblock {\em Journal of Mathematical Fluid Mechanics}, 25(2):28, 2023.

\bibitem{Chen-Kraichnan-1989}
{\ifshowrevisions\color{red}\fi
Shiyi Chen and Robert~H. Kraichnan.
\newblock Sweeping decorrelation in isotropic turbulence.
\newblock {\em Physics of Fluids A: Fluid Dynamics}, 1(12):2019--2024, 1989.
}

\bibitem{cotter2017stochastic}
Colin~J Cotter, Georg~A Gottwald, and Darryl~D Holm.
\newblock Stochastic partial differential fluid equations as a diffusive limit
  of deterministic {L}agrangian multi-time dynamics.
\newblock {\em Proceedings of the Royal Society A: Mathematical, Physical and
  Engineering Sciences}, 473(2205):20170388, 2017.

\bibitem{Craig-Sulem93}
Walter Craig and Catherine Sulem.
\newblock Numerical simulation of gravity waves.
\newblock {\em Journal of Computational Physics}, 108(1):73--83, 1993.

\bibitem{crisan2019solution}
Dan Crisan, Franco Flandoli, and Darryl~D Holm.
\newblock Solution properties of a 3{D} stochastic {E}uler fluid equation.
\newblock {\em Journal of Nonlinear Science}, 29(3):813--870, 2019.

\bibitem{Debusshe-Hug-Memin-2023}
Arnaud Debussche, Berenger Hug, and Etienne M{\'e}min.
\newblock A consistent stochastic large-scale representation of the
  {N}avier--{S}tokes equations.
\newblock {\em Journal of Mathematical Fluid Mechanics}, 25(1):19, 2023.

\bibitem{Debussche-Memin25}
Arnaud Debussche and Etienne M\'{e}min.
\newblock Variational principles for fully coupled stochastic fluid dynamics
  across scales.
\newblock {\em Physica D: Nonlinear Phenomena}, 481:134777, 2025.

\bibitem{debussche2024derivation}
Arnaud Debussche, Etienne M{\'e}min, and Antoine Moneyron.
\newblock Derivation of stochastic models for coastal waves.
\newblock {\em Stochastic Transport in Upper Ocean Dynamics III}, pages
  183--222, 2024.

\bibitem{Moneyron-2025}
Arnaud Debussche, Etienne M{\'e}min, and Antoine Moneyron.
\newblock Some properties of a non-hydrostatic stochastic oceanic primitive
  equations model.
\newblock {\em ArXiv}, 2025.

\bibitem{Debussche-Pappalattera2023}
Arnaud Debussche and Umberto Pappalettera.
\newblock Second order perturbation theory of two-scale systems in fluid
  dynamics.
\newblock {\em J. Eur. Math. Soc.}, 28(4):1533--1595, 2026.

\bibitem{dinvay2022}
Evgueni Dinvay and Etienne M{\'e}min.
\newblock {Hamiltonian formulation of the stochastic surface wave problem}.
\newblock {\em {Proceedings of the Royal Society A}}, pages 1--26, September
  2022.

\bibitem{Ferrari-Wunsch-09}
Raffaele Ferrari and Carl Wunsch.
\newblock Ocean circulation kinetic energy: Reservoirs, sources, and sinks.
\newblock {\em Annual Review of Fluid Mechanics}, 41(Volume 41, 2009):253--282,
  2009.

\bibitem{Flandoli-Galeati-Luo-2021}
Franco Flandoli, Lucio Galeati, and Dejun Luo.
\newblock Delayed blow-up by transport noise.
\newblock {\em Communications in Partial Differential Equations},
  46(9):1757--1788, 2021.

\bibitem{Flandoli-Luo-2021}
Franco Flandoli and Dejun Luo.
\newblock High mode transport noise improves vorticity blow-up control in 3{D}
  {N}avier--{S}tokes equations.
\newblock {\em Probability Theory and Related Fields}, 180(1):309--363, 2021.

\bibitem{Flandoli-Pappalettera-2023}
Franco Flandoli and Umberto Pappalettera.
\newblock 2{D} {E}uler equations with stratonovich transport noise as a
  large-scale stochastic model reduction.
\newblock {\em Journal of Nonlinear Science}, 31(1):24, 2021.

\bibitem{Flandoli-Russo2023}
Franco Flandoli and Francesco Russo.
\newblock Reduced dissipation effect in stochastic transport by {G}aussian
  noise with regularity greater than 1/2.
\newblock {\em ArXiv}, 2305.19293, 2023.

\bibitem{Frankignoul-Hasselmann-1977}
Claude Frankignoul and Klaus Hasselmann.
\newblock Stochastic climate models. part {II}. application to sea-surface
  temperature anomalies and thermocline variability.
\newblock {\em Tellus}, 29:289--305, 1977.

\bibitem{Galeati-Luo-2020}
Lucio Galeati.
\newblock On the convergence of stochastic transport equations to a
  deterministic parabolic one.
\newblock {\em Stochastics and Partial Differential Equations: Analysis and
  Computations}, 8(4):833--868, 2020.

\bibitem{Galeati-Luo2023}
Lucio Galeati and Dejun Luo.
\newblock Weak well-posedness by transport noise for a class of 2d fluid
  dynamics equations.
\newblock {\em Journal of Functional Analysis}, 289(12):111158, 2025.

\bibitem{goda2000}
Yoshimi Goda.
\newblock {\em Random Seas and Design of Maritime Structures}.
\newblock World Scientific, Singapore, 2nd edition, 2000.

\bibitem{Goodair-et-al2022}
Daniel Goodair, Dan Crisan, and Oana Lang.
\newblock Existence and uniqueness of maximal solutions to spdes with
  applications to viscous fluid equations.
\newblock {\em Stochastics and Partial Differential Equations: Analysis and
  Computations}, 12:1201--1264, 2024.

\bibitem{hasselmann1962nonlinear}
Klaus Hasselmann.
\newblock On the non-linear energy transfer in a gravity-wave spectrum. part 1.
  general theory.
\newblock {\em Journal of Fluid Mechanics}, 12(4):481--500, 1962.

\bibitem{hasselmann1963nonlinear}
Klaus Hasselmann.
\newblock On the non-linear energy transfer in a gravity wave spectrum part 2.
  conservation theorems; wave-particle analogy; irreversibility.
\newblock {\em Journal of Fluid Mechanics}, 15(2):273--281, 1963.

\bibitem{Hasselmann-1976}
{\ifshowrevisions\color{red}\fi
Klaus Hasselmann.
\newblock Stochastic climate models part {I.} theory.
\newblock {\em Tellus}, 28(6):473--485, 1976.
}

\bibitem{hasselmann1973}
Klaus Hasselmann, Tim~P. Barnett, Evert Bouws, H.~Carlson, David~E. Cartwright,
  Klaus Enke, J.~A. Ewing, Hans Gienapp, Dieter~E. Hasselmann, P.~Kruseman,
  A.~Meerburg, Peter Muller, Dirk~J. Olbers, Klaus Richter, Werner Sell, and
  Hans Walden.
\newblock Measurements of wind-wave growth and swell decay during the joint
  north sea wave project (jonswap).
\newblock {\em Deutsche Hydrographische Zeitschrift}, 8(12):1--95, 1973.

\bibitem{hasselmann1976}
Klaus Hasselmann, Duncan~B. Ross, Peter Muller, and Werner Sell.
\newblock A parametric wave prediction model.
\newblock {\em Journal of Physical Oceanography}, 6(2):200--228, 1976.

\bibitem{Holm2015}
Darryl~D. Holm.
\newblock Variational principles for stochastic fluid dynamics.
\newblock {\em Proceedings of the Royal Society A}, 471(2176):20140963, 2015.

\bibitem{komen1994}
Gerbrand~J. Komen, Luigi Cavaleri, Mark Donelan, Klaus Hasselmann, Susanne
  Hasselmann, and Peter A. E.~M. Janssen.
\newblock {\em Dynamics and Modelling of Ocean Waves}.
\newblock Cambridge University Press, Cambridge, 1994.

\bibitem{Kraichnan-1964}
{\ifshowrevisions\color{red}\fi
Robert~H. Kraichnan.
\newblock Kolmogorov's hypotheses and {E}ulerian turbulence theory.
\newblock {\em The Physics of Fluids}, 7(11):1723--1734, 1964.
}

\bibitem{Lang2023}
Oana Lang, Dan Crisan, and {\'E}tienne M{\'e}min.
\newblock Analytical properties for a stochastic rotating shallow water model
  under location uncertainty.
\newblock {\em Journal of Mathematical Fluid Mechanics}, 25(2):29, 2023.

\bibitem{Lannes-2013}
David Lannes.
\newblock {\em The water waves problem: mathematical analysis and asymptotics},
  volume 188.
\newblock American Mathematical Soc., 2013.

\bibitem{Li-et-al-JPO-25}
Long Li, Etienne M{\'e}min, and Bertrand Chapron.
\newblock A generalized stochastic formulation of the ekman-stokes model with
  statistical analyses.
\newblock {\em Journal of Physical Oceanography}, 55(9):1389 -- 1407, 2025.

\bibitem{Luke67}
Jon~Christian Luke.
\newblock A variational principle for a fluid with a free surface.
\newblock {\em Journal of Fluid Mechanics}, 27(2):395--397, 1967.

\bibitem{Majda-et-al-1999}
{\ifshowrevisions\color{red}\fi
Andrew~J. Majda, Ilya Timofeyev, and Eric Vanden~Eijnden.
\newblock Models for stochastic climate prediction.
\newblock {\em Proceedings of the National Academy of Sciences},
  96(26):14687--14691, 1999.
}

\bibitem{Majda-et-al-2001}
{\ifshowrevisions\color{red}\fi
Andrew~J. Majda, Ilya Timofeyev, and Eric Vanden~Eijnden.
\newblock A mathematical framework for stochastic climate models.
\newblock {\em Communications on Pure and Applied Mathematics}, 54(8):891--974,
  2001.
}

\bibitem{McWilliams2016}
James~C. McWilliams.
\newblock Submesoscale currents in the ocean.
\newblock {\em Proceedings of the Royal Society A}, 472(2189):20160117, 2016.

\bibitem{McWilliams-Huckle-2006}
James~C. McWilliams and Edward Huckle.
\newblock Ekman layer rectification.
\newblock {\em Journal of Physical Oceanography}, 36(8):1646--1659, 2006.

\bibitem{memin2014fluid}
Etienne M{\'e}min.
\newblock Fluid flow dynamics under location uncertainty.
\newblock {\em Geophysical \& Astrophysical Fluid Dynamics}, 108(2):119--146,
  2014.

\bibitem{DebusscheMemin2026b}
{\ifshowrevisions\color{red}\fi
Etienne M{\'e}min, Bertrand Chapron, Arnaud Debussche, and Louis Mari\'e.
\newblock Stochastic transport and wave interactions for multiscale surface
  gravity waves: Part ii: Kinetic theory and ocean-wave applications.
\newblock {\em ArXiv}, 2026.
\newblock arXiv:2608.27105; companion paper II.
}

\bibitem{Miles1977}
{\ifshowrevisions\color{red}\fi
John~W. Miles.
\newblock On {Hamilton}'s principle for surface waves.
\newblock {\em Journal of Fluid Mechanics}, 83(1):153--158, 1977.
}

\bibitem{PavliotisStuart2008}
{\ifshowrevisions\color{red}\fi
Grigorios~A. Pavliotis and Andrew~M. Stuart.
\newblock {\em Multiscale Methods: Averaging and Homogenization}, volume~53 of
  {\em Texts in Applied Mathematics}.
\newblock Springer, New York, 2008.
}

\bibitem{Polnikov2004}
Vladislav~G. Polnikov.
\newblock On the problem of numerical calculation of the {Hasselmann} kinetic
  integral.
\newblock {\em Oceanology}, 44(1):1--10, 2004.

\bibitem{Resseguier2017a}
Valentin Resseguier, Etienne M{\'e}min, and Bertrand Chapron.
\newblock Geophysical flows under location uncertainty. {Part I}: Random
  transport and general models.
\newblock {\em Geophysical \& Astrophysical Fluid Dynamics}, 111(3):149--176,
  2017.

\bibitem{Resseguier2017b}
Valentin Resseguier, Etienne M{\'e}min, and Bertrand Chapron.
\newblock Geophysical flows under location uncertainty. {Part II}:
  Quasi-geostrophy and efficient ensemble prediction.
\newblock {\em Geophysical \& Astrophysical Fluid Dynamics}, 111(3):177--206,
  2017.

\bibitem{Saravanan-McWiliams-1998}
Ramalingam Saravanan and James~C. McWilliams.
\newblock Advective ocean--atmosphere interaction: An analytical stochastic
  model with implications for decadal variability.
\newblock {\em Journal of Climate}, 11(2):165--188, 1998.

\bibitem{Street-Crisan-23}
Oliver Street and Dan Crisan.
\newblock Semi-martingale driven variational principles.
\newblock {\em Proceedings of the Royal Society A: Mathematical, Physical and
  Engineering Sciences}, 477(2247):20200957, 2021.

\bibitem{Street-2023}
Oliver~D. Street.
\newblock A structure preserving stochastic perturbation of classical water
  wave theory.
\newblock {\em Physica D: Nonlinear Phenomena}, 447:133689, 2023.

\bibitem{Sullivan2010DynamicsWinds}
Peter~P. Sullivan and James~C. McWilliams.
\newblock Dynamics of winds and currents coupled to surface waves.
\newblock {\em Annual Review of Fluid Mechanics}, 42:19--42, 2010.

\bibitem{Tennekes-1975}
Henk Tennekes.
\newblock Eulerian and {L}agrangian time microscales in isotropic turbulence.
\newblock {\em Journal of Fluid Mechanics}, 67(3):561--567, 1975.

\bibitem{WAMDI1988}
{The WAMDI Group}.
\newblock The {WAM} model---a third generation ocean wave prediction model.
\newblock {\em Journal of Physical Oceanography}, 18(12):1775--1810, 1988.

\bibitem{Thomas2020}
Justin~A. Thomas, Oliver B\"uhler, and K.~Shafer Smith.
\newblock Internal waves and their interactions with surface waves.
\newblock {\em Annual Review of Marine Science}, 12:1--25, 2020.

\bibitem{Tolman2009}
Hendrik~L. Tolman.
\newblock User manual and system documentation of {WAVEWATCH III} version 3.14.
\newblock Technical Report 276, NOAA/NWS/NCEP/MMAB, Camp Springs, MD, 2009.

\bibitem{Tucciarone-et-al-James-25}
Francesco~L. Tucciarone, Long Li, Etienne M{\'e}min, and Pranav Chandramouli.
\newblock Derivation and numerical assessment of a stochastic large-scale
  hydrostatic primitive equations model.
\newblock {\em Journal of Advances in Modeling Earth Systems},
  17(7):e2024MS004783, 2025.
\newblock e2024MS004783 2024MS004783.

\bibitem{VillasBoas2020WaveCurrent}
Ana~B. Villas~B{\^o}as, Bruce~D. Cornuelle, Matthew~R. Mazloff, Sarah~T. Gille,
  and Fabrice Ardhuin.
\newblock Wave-current interactions at meso- and submesoscales: Insights from
  idealized numerical simulations.
\newblock {\em Journal of Physical Oceanography}, 50(12):3483--3500, 2020.

\bibitem{young1999}
Ian~R. Young.
\newblock {\em Wind Generated Ocean Waves}.
\newblock Elsevier, Amsterdam, 1999.

\bibitem{Zakharov68}
Vladimir~E. Zakharov.
\newblock Stability of periodic waves of finite amplitude on the surface of a
  deep fluid.
\newblock {\em Journal of Applied Mechanics and Technical Physics},
  9(2):190--194, 1968.

\end{thebibliography}
\endgroup
\end{document}